%% file: main.tex
\documentclass[preprint,12pt,3p]{elsarticle}
\usepackage{amssymb}
\usepackage{comment}
\usepackage{times}
\usepackage{latexsym}
\usepackage[neverdecrease]{paralist}
\usepackage{subcaption}
\usepackage{graphicx}
\usepackage{booktabs}
\usepackage[table]{xcolor}
\usepackage{graphics}
\usepackage{tabularx}
\usepackage{xspace}
\usepackage{graphicx}
\usepackage{booktabs}
\usepackage{mdwlist}
\usepackage{multirow}
\usepackage{amsmath}
\usepackage{enumitem}
\usepackage{calc}
\usepackage{url}
\usepackage[ruled,linesnumbered]{algorithm2e}
\usepackage{booktabs}
\usepackage{color}
\usepackage{color,soul}

\usepackage{amssymb}% http://ctan.org/pkg/amssymb
\usepackage{pifont}% http://ctan.org/pkg/pifont

\makeatletter
\def\BState{\State\hskip-\ALG@thistlm}
\makeatother

\newcommand{\urlwofont}[1]{ \urlstyle{same}\url{#1} }

\newcommand{\nores}{
\cellcolor{gray} & \cellcolor{gray} & \cellcolor{gray} & \cellcolor{gray}
}

\renewcommand{\vec}[1]{\mathbf{#1}}

\newcommand{\NMISQRT}{$\mathrm{\sqrt{NMI}}$}
\newcommand{\NMIMAX}{NMI}

\newcommand{\Combinatorial}{Comb.}
\newcommand{\Mnv}{Mn2v}
\newcommand{\CAnv}{Cn2v}
\newcommand{\MCAnv}{MCn2v}

\newcommand{\para}[1]{\paragraph{\textnormal{\textbf{#1}}}}

\begin{document}

\begin{frontmatter}

%\title{Community Preserving Node Embedding}
%\title{Embedding-based Network Community Detection with Combinatorial Heuristics and Maximum-Entropy based Walk}
%\title{A Combination of Combinatorial Heuristics and Maximum-Entropy based Walk for Community-Structure aware Node Embedding}
\title{Community Structure aware Embedding of Nodes in a Network}

\author[label1]{Swarup Chattopadhyay}
%\corref{cor1}\fnref{label3}
\address[label1]{Indian Statistical Institute, Kolkata, India}
\ead{swarup_r@isical.ac.in}

\author[label2]{Debasis Ganguly}
%\corref{cor1}\fnref{label3}
\address[label2]{IBM Research, Dublin, Ireland}
\ead{debasis.ganguly1.ie.ibm.com}
%\ead[url]{author-one-homepage.com}

%\cortext[cor1]{I am corresponding author}
%\fntext[label3]{I also want to inform about\ldots}
%\fntext[label4]{Small city}

%\author[label5]{Author Two}
%\address[label5]{Some University}
%\ead{author.two@mail.com}

%\author[label1,label5]{Author Three}
%\ead{author.three@mail.com}

\begin{abstract}
\input{abstract.tex}
\end{abstract}


\end{frontmatter}

%%
%% Start line numbering here if you want
%%
% \linenumbers

\input{revision-v1/content}

\bibliographystyle{elsarticle-num}
\bibliography{main.bib}

\end{document}

%% file: abstract.tex
Detecting communities or the modular structure of real-life networks (e.g. a social network or a product purchase network) is an important task because the way a network functions is often determined by its communities.
Traditional approaches to community detection involve modularity-based algorithms, which generally speaking, construct partitions based on heuristics that seek to maximize the ratio of the edges within the partitions to those between them. On the other hand, node embedding approaches represent each node in a graph as a real-valued vector and is thereby able to transform the problem of community detection in a graph to that of clustering a set of vectors. Existing node embedding approaches are primarily based on, first, initiating random walks from each node to construct a context of a node, and then make the vector representation of a node close to its context. However, standard node embedding approaches do not directly take into account the community structure of a network while constructing the context around each node.
To alleviate this, we explore two different threads of work. First, we investigate the use of maximum entropy based random walks to obtain more centrality preserving embedding of nodes, which may lead to more effective clusters in the embedded space.
Second, we propose a community structure aware node embedding approach, where we incorporate modularity-based partitioning heuristics into the objective function of node embedding.
We demonstrate that our proposed combination of the combinatorial and the embedding approaches for community detection outperforms a number of modularity-based baselines, and K-means clustering on a standard node-embedded (node2vec) vector space
on a wide range of real-life and synthetic networks of different sizes and densities.
%Extensive experiments have also been performed by varying parameters of node2vec. The empirical results suggest the effectiveness of the proposed method compared to node2vec (and other baselines) for a wide range of these parameter settings.

%% file: keywords.tex
Community Detection,
Node Embedding,
Combinatorial Approaches,
K-means Clustering,
Maximum Entropy Random walk

%% file: revision-v1/content.tex
\section{Introduction}

A \emph{network community} represents a set of nodes with a relatively dense set of connections between its members and relatively sparse connections between
its member nodes and the ones outside the community.
Partitioning a network (graph) into communities usually leads to better analyzing the functionality of the network and is of immense practical interest for real-world networks, because such communities potentially represent organizational units in social networks \cite{girvan2002community}, scientific disciplines in authorship-citation academic publications networks \cite{Newman2004modularity1}, or functional units in biological networks (e.g. protein-protein interactions) \cite{fgene}. 

Traditional approaches of community detection incrementally construct a community (set of nodes) by employing an objective function \cite{Newman2004modularity1} that seeks to maximize its internal connectivity and minimize the number of external edges \cite{newman2006finding}. Recent research also focused on high quality, scalable and parallel community detection for large real graphs by maximizing a metric based on triangle analysis \cite{prat2014high}. Modularity \cite{Newman2004modularity1} is a widely used goodness metric that effectively measures the strength of the community structure of a network. An algorithm for inferring community structure from a large network by
greedily optimizing the modularity value of a network is proposed in \cite{clauset2004finding}. The Louvain algorithm is a greedy approach that produces a hierarchy of communities by maximizing the modularity value \cite{blondel2008fast}. A multilevel refinement based variant of the Louvain algorithm was proposed in \cite{waltman2013smart}. Apart from the modularity based approach, several other heuristic approaches, such as those based on information theoretic principles \cite{rosvall2008maps}, or those based on random walks \cite{raghavan2007near}, have also been proposed in order to detect both the disjoint and overlapping communities of a network. These heuristic approaches often fail to preserve the locality information of a node, thereby resulting in ineffective community detection.

Node embedding based approaches, on the other hand, first transform a graph from a discrete representation to a continuous one (where each node is represented by a vector of real numbers) and then clusters the space of the embedded node vectors to predict the communities.
Specifically, a graph representation learning approach, such as Deepwalk \cite{Perozzi:2014} and Node2vec \cite{Grover:2016} represents each node of a graph as a real-valued vector seeking to preserve the correlation between the topological properties of the discrete graph with the distance measures in the embedded metric space. For example, the vectors corresponding to a pair of nodes in the embedded space is usually close (low distance or high inner product similarity), if it is likely to visit one of the node of the pair with a random walk initiated from the other one.
However, a major limitation of the random walk based node representation approach is that a random walk may span across the community from which it stared with, which eventually could lead to representing nodes from different communities in close proximity in an embedded space. This in turn can may not result in effective community detection on application of a standard clustering algorithm, e.g. K-means, in the embedded space of nodes.

Ideally speaking, for effective community detection with a clustering algorithm operating on the embedded space of node vectors, a node embedding algorithm should preserve the community structure from the discrete space of the sets of nodes to the continuous space of real-valued vectors as perceived with the conventional definitions of the distance metric (e.g. $l_2$ distance) and the inner product between pairs of vectors denoting the similarity between them.
In other words, a central (hub) node of a community in the discrete graph representation should be transformed in the embedded space in such a way so that it contains other vectors, corresponding to the nodes of the other members in the community, in its close neighborhood. In our study, we investigate two methods to achieve such a transformation.

\para{\textbf{Our Contributions}}
To alleviate the problems of combinatorial and node embedding approaches, we investigate if a combination of the two can overcome the individual limitations.
%In particular, the specific contributions of this paper are as follows.
%The main purpose here is how node embedding in combination with combinatorial approaches helps in better identification of the actual communities in a network. Therefore the main contribution of this article is two folded:
%\begin{itemize}

%\item
Our first contribution seeks to address the local effects of the random walk of standard node embedding, namely node2vec \cite{Grover:2016}, or DeepWalk \cite{Perozzi:2014}.
Specifically, in contrast to the first-order and second-order random walks based contextualization of nodes in DeepWalk \cite{Perozzi:2014} and node2vec \cite{Grover:2016}, respectively, we investigate a maximum-entropy biased random walk (MERW) \cite{maxentropywalk}, in which the transition probabilities are non-local, i.e., they depend on the structure of the entire graph rather than on the very local neighborhood of a node.
  
%\item
Second, we investigate if traditional approaches to community detection that operate on a discrete graph (adjacency matrix), e.g. modularity-heuristic \cite{clauset2004finding} or InfoMap \cite{rosvall2008maps}, Label Propagation Algorithm (LPA) \cite{raghavan2007near}, can be leveraged to better contextualize a node for the purpose of obtaining its embedded representation. In other words, while training a classifier that learns to predict a node vector's context, we favour those cases where the context nodes are likely to be a part of the same community as that of the current node; these likelihoods being estimated with a modularity-based heuristic.

Additionally, we also investigate a combination of the two different community aware embedding approaches, i.e. employing MERW to first contextualize the nodes and then using the preferential training based on the modularity heuristic.

%\end{itemize}

The remainder of the paper is organized as follows. Section \ref{sec:relwork} provides a background and brief overview of some related works. Our proposed method of node embedding is explained in Section \ref{sec:methodology}. Section \ref{sec:experimentsetup} provides details on the setup of our experiments. Section \ref{sec:results} presents the results and their analysis over several real world and synthetic networks. Finally, Section \ref{sec:conclusions} concludes the paper with directions of future research.

\section{Background Review and Related Work}   \label{sec:relwork}

\para{Review of Combinatorial Approaches}

We review a number of combinatorial approaches to community detection. Each combinatorial approach has the common underlying principle of first constructing an initial partition of an input graph into a set of sub-graphs (communities) and then refining the partition at every iterative step. Among a number of possible ways to modify a current partition, the one that
maximizes a global objective function is chosen. The global objective, in turn, is computed by aggregating the local objectives over and across the constituent sub-graphs.

\emph{Modularity} is defined as an intrinsic measure of how effectively, with respect to its topology, a graph (network) is partitioned into a given set of communities \cite{clauset2004finding}. More formally, given a partition of a graph $G=(V,E)$ into $p$ communities, i.e. given an assigned community (label) $c_v \in \{1,\ldots,p\}$ for each node $v \in V$, the modularity, $Q$ is defined as the expected ratio of the number of intra-community edges to the total number of edges, the expectation being computed with respect to the random case of
assigning the nodes to arbitrary communities.
More specifically,
\begin{equation}
Q=\frac{1}{2|E|}\sum_{vw}\big(A_{vw}-\frac{k_vk_w}{2|E|}\mathbb{I}(c_v=c_w)\big) \label{eq:modularity},
\end{equation}
where $A_{vw}$ denotes the adjacency relation between nodes $v$ and $w$, i.e. $A_{vw}=1$ if $(v,w) \in E$;
$k_v$ denotes the number of edges incident on a node $v$; $\mathbb{I}(c_v,c_w)$ indicates if nodes $v$ and $w$ are a part of the same community. A high value of $Q$ in Equation \ref{eq:modularity} represents
a substantial deviation of the 
fraction of intra-community edges to the total number of edges from what one would expect for a randomized network. The study in \cite{clauset2004finding} suggests that
a value above $0.3$ is often a good indicator of significant community structure in a network.

The `CNM' (Clauset Newman Moore) algorithm
\cite{Newman2004modularity1}
proposes a greedy approach that seeks to optimise the modularity score (Equation \ref{eq:modularity}). Concretely speaking, it starts with an initial state of node being assigned to a distinct singleton community, seeking to refine the current assignment at every iteration by merging a pair of communities that yields the maximum improvement of the modularity score.
The algorithm proceeds until it is impossible to find a pair of communities which if merged yields an improvement in the modularity score.

The `Louvain' or the `Multilevel' algorithm \cite{blondel2008fast} involves first greedily assigning nodes to communities, favoring local optimizations of modularity, and then
repeating the algorithm on a coarser network constructed from the communities found in the first step. These two steps are repeated until no further modularity increasing reassignments are found.

`SCDA' (Scalable Community Detection Algorithm) \cite{prat2014high}
detects disjoint communities in networks by maximizing WCC, a recently proposed community metric \cite{prat2012shaping} based on
triangle structures within a community. SCD implements a two-phase procedure that combines different strategies. In the first phase, SCD uses the clustering coefficient as an heuristic to obtain a preliminary partition of the graph. In the second phase, SCD refines the initial partition by moving vertices between communities as long as the WCC of the communities increase.

The work in \cite{jiang2010spici} proposed a scalable algorithm - `SPICi' (`Speed and Performance In Clustering' and pronounced as `spicy'), which constructs communities of nodes by first greedily starting from local seed sets of nodes with high degrees, and then adding those nodes to a cluster that maximize a two-fold objective of
the density and the adjacency of nodes within the cluster. The underlying principle of SPICi is similar to that of `DPClus' \cite{altaf2006development}, the key differences being SPICi exploits a simpler cluster expansion approach, uses a different seed selection criterion and incorporates interaction confidences.

The `LEADE' (Leading Eigenvector) method applies a spectral decomposition of the modularity matrix $M$, defined as
%\begin{equation}
$M_{vw} = A_{vw} - \frac{k_vk_w}{2|E|}$ \cite{newman2006finding}, where %  
%\end{equation}
the leading eigenvector is used to split the graph into two sub-graphs for maximizing the modularity score. The process is then recursively applied on each sub-graph until the modularity value cannot be improved further.
%Its computational complexity of each graph bipartition is $O ( N ( E + N )) $, or $O ( N^{2})$ on a sparse graph.
%
`LPA' (Label Propagation Algorithm) \cite{raghavan2007near} relies on the assumption
that each node of a network is assigned to the same community as the majority of its neighbours. The algorithm starts with initialising a distinct label (community) for each node in the network. Each 
node, visited in a random order, then takes the label of the majority of its neighbours. The iteration stops when the label assignments cannot be changed further. 
%
%The computational complexity of label propagation algorithm is $ O ( E ) $.

The `InfoMap' algorithm \cite{rosvall2008maps} finds the optimal encoding of a network based on maximizing the information needed to compress the movement of a random walker across communities on the one hand, whereas 
minimizing the code length to represent this information. The algorithm makes uses of the core idea that a random walk initiated from a node that is central to a community is less likely to visit a node of a different community. Huffman encoding of such nodes, hence, are likely to be shorter. 
%This algorithm runs in $ O( E )$.
%
The `WalkTrap' algorithm \cite{pons2005computing}
is a hierarchical agglomerating clustering (HAC) algorithm using an idea similar to InfoMap that short length random walks tend to visit only the nodes within a single community. The distance metric that the algorithm uses for the purpose of HAC between two sets of nodes is the distance between the probability distributions of nodes visited by random walks initiated from member nodes of the two sets.  

Different from the existing work in combinatorial approaches to community detection, in our work, we propose a framework to integrate a combinatorial approach within the framework of an embedding approach (specifically, node2vec).

\para{Review of Embedding Approaches}

In contrast to the combinatorial approaches which directly work on the discrete space (vertices and edges) of a graph, $G=(V,E)$, an embedding approach transforms each node of a graph, $u$, into a real-valued vector, $\vec{u}$, seeking to preserve the topological structure of the nodes. Formally,
%\begin{equation}
$\theta: u \mapsto \vec{u} \in \mathbb{R}^d,\,\, \forall u \in V$, %\label{eq:transform}.
%\end{equation}
where the transformation function $\theta$ is learned with the help of noise contrastive estimation, i.e., the objective is to make the similarity (inner product) between vectors for nodes $u$ and $v$ higher if $v$ lies in the neighborhood of $u$, and to be of a value small if $v$ does not belong to the neighborhood of $u$ (e.g. $v$ being a randomly sampled node from the graph).
Formally, 
\begin{equation}
J(\theta) = \sum_{u} \sum_{v \in \mathcal{N}(u)}P(y=1|\vec{u},\vec{v}) +
\sum_{u} \sum_{\bar{v} \in \bar{\mathcal{N}}(u)}P(y=0|\vec{u},\vec{\bar{v}}) \label{eq:node2vec},
\end{equation}
where $y$ denotes a binary response variable to train the likelihood function, where $\mathcal{N}(u)$ denotes the neighborhood of node $u$, and the negative component ($y=0$) in the likelihood function refers to the randomly sampled noise (the number of negative samples is determined by the way the complement of the neighborhood, $\bar{\mathcal{N}}$, is defined). 

Popular approaches to learn the transformation function, $\theta$, of Equation \ref{eq:node2vec} includes node2vec \cite{Grover:2016} and DeepWalk \cite{Perozzi:2014}, which differ in the way the neighborhood function, $\mathcal{N}(u)$, is defined. While DeepWalk uses a uniform random walk to constitute the neighborhood or context of a node, node2vec uses a biased random walk (with a relative importance to depth-first or breadth-first traversals).

A transformation of the nodes as real-valued vectors then allows the application of relatively simple (but effective) clustering approaches, such as K-means, to partition the embedding space of nodes into distinct clusters. This is because in contrast to the discrete space, the vector space is equipped with a metric function which allows to compute distance (or equivalently similarity) between \emph{any} pair of nodes (as opposed to the discrete case).

The study in \cite{Cavallari:2017} proposed an expectation-maximization (EM) based approach to iteratively refine a current community assignment (initialized randomly) using node embeddings. The objective was to ensure that the embedded vectors of each community fits a Gaussian mixture model, or in other words, the embedded space results in relatively disjoint convex clusters. The main difference of our approach with that of \cite{Cavallari:2017} is that while \cite{Cavallari:2017}
uses additional terms in the objective of function of node embedding which seeks to maximize the similarity of a node vector with its cluster \emph{centroid}, our proposed model works only with node pairs and does explicitly make use of the centroid vectors.

The work in \cite{Wang:2016} proposed to include an additional term in the objective of the transformation function (Equation \ref{eq:node2vec}) corresponding to the \emph{second order similarity} between the neighborhoods of two nodes. Different to \cite{Wang:2016}, which seeks to obtain a general purpose embedding of graphs, we rather focus only on the community detection problem. 

%Furthermore, Maximal-Entropy Random Walk \cite{Ochab2013} have used to obtain more accurate neighbourhood information's compared to the uniform random walk and combine it with the node2vec objective function for better identification of the actual communities present in a network. 

Different from random walk based node embedding, other forms of embedded representations of graphs include those of
applying depth-based convolutional autoencoders \cite{conv_auto_encoder}, and its quantum theory based extension \cite{quantum}. Similar to the neighborhood based representation learning of graph nodes in an Euclidean space, such neighborhood based contrastive learning also finds application for discriminative feature extraction, e.g. the work in \cite{isomap}
minimizes pairwise intra-class distances in the same manifold and
maximizes the inter-class ones between different manifolds.

\section{Proposed Methodology}  \label{sec:methodology}

We start this section by first discussing the potential problems of applying K-means clustering on the output of the Node2vec \cite{Grover:2016} embedding approach. We follow it up with our proposed changes in the objective function of a node embedding algorithm that better preserves the topological structure of the communities (from the discrete space of vertices and their relations) to their embedded representations in a Cartesian space of reals. 

\subsection{Limitations of Node2vec for Community Detection}

Random walk over a network usually helps in getting better contextualization of a particular node through exploring its nearest neighbours in order to detect communities in a network. A random walk of length $l$ on a connected graph $G=(V,E)$, started at vertex $u$, is a random sequence $(u, u_{1}, ... ,u_{l})$ of vertices, such that neighbors in the sequence are connected in $G$. A random walk, such as the one used in node2vec \cite{Grover:2016} and DeepWalk \cite{Perozzi:2014}, is usually \emph{local} in nature, meaning a choice of the next node to visit in sequence depends only on the current node itself. 

%Generally, a walker repeatedly moves along an edge to a selected neighbor using transition probability, until $l$ edges have been crossed.
%Thus the random walk can be easily biased by assigning static edge weight as the transition probability between the nodes while sampling the next node.
Specifically, if $P \in \mathbb{R}^{|V|\times|V|}$ denotes the stochastic transition matrix of a graph $G=(V,E)$, where $P_{uv}$ denotes the probability of visiting node $v$ in sequence after visiting node $u$, in a standard uniform random walk (URW), this probability is given by
\begin{equation}
P_{uv} = \frac{A_{uv}}{k_u},\,\, k_u = |\{w:(u,w) \in V\}| \label{eq:urw},
\end{equation}
where $k_u$ denotes the degree of node $u$. In other words, Equation \ref{eq:urw} indicates that there is an equal likelihood of choosing a node $v$ as the next node in sequence from the neighbors of node $u$.
Note that if there are no edges between $u$ and $v$ (i.e. $A_{uv}=0$), $v$ cannot be visited after $u$, i.e. $P_{uv}=0$. 

The random walk used in node2vec \cite{Grover:2016} introduces a bias to this uniform walk (of Equation \ref{eq:urw}) to relatively control the spread of the walk.
Specifically, node2vec uses two parameters - $p$ (return parameter) and $q$ (in-out parameter), which control how likely it is for the walk to stay close to its starting point (akin to breadth-first search or BFS), or how likely it is for the walk to visit nodes with relatively high hop-counts with respect to the initial node (akin to depth-first search or DFS). It is reported in \cite{Grover:2016} that low values of $q$ in combination with high values of $p$ typically favours a higher exploration of the graph in a DFS manner, whereas high values of $q$ in combination with low values of $p$ is likely to constrain the walk locally.

%The effectiveness of the return and In-out parameters in terms of community discovery is discussed in the section 6.2 in details.

A point to note is that since different choices of $p$ and $q$ lead to different walks (contexts around a node), they are also likely lead to different vector representations of the nodes of a network in an embedded space. The different relative distances between the node vectors 
is, in turn, likely to affect the clustering effectiveness in the embedded space (and that of detecting the true communities of the original graph).

%which may affect the performances of any computational approaches for further structural analysis of a real world network. In particular, for community discovery, it may affect in preserving the local community structure of a network in the embedded space.

\input{plot_of_main_ideas/plot_main_idea.tex}

Concretely speaking, the cases where a node belongs to the periphery of a (true) community are the ones that are likely to lead to introducing false positives (in the form of a node from another community) in the context of the node thus failing to preserve the structural equivalence hypothesis of the discrete graph with the embedded vector space \cite{henderson2012rolx}.
The failure to preserve the structural equivalence could lead to a small distance between two vectors belonging to two different (true) communities in the embedded space, which could in turn lead to falsely including the vectors of these two nodes in a single cluster during the clustering step.

The idea is illustrated in Figure \ref{fig:n2v-schematic}, which shows how random walks spanning across two different communities can affect the relative distances between the embedded vectors in the embedded space, which in turn can degrade the quality of K-means clustering based community detection. 

%The random walk used in \cite{Grover:2016}, irrespective of its Breadth-first Sampling (BFS) or Depth-first Sampling (DFS), fails to the proper vector representations of nodes which belong to the outer part of a community or belongs to a overlapped part between two communities. More preciously, the work presented in \cite{Grover:2016} fails to represents/contextualize a node which belongs to a periphery of a community in a network and hence fails to meet the structural equivalence hypothesis \cite{henderson2012rolx} which suggests that the nodes having similar structural roles should be embedded closely together in a network. 

Our proposed embedding algorithm seeks to alleviate this problem in two ways. First, we make use of a biased random walk (specifically maximum entropy-based walk), which leverages structural information at a \emph{global} level (instead of selecting the next node to visit in a walk on the basis of the edge weights of its neighbors only \cite{maxentropywalk}). Second, we make the embedding objective function aware of an initial estimate about the community structure of a network on the basis of a combinatorial approach, e.g. the modularity criterion \cite{Newman2004modularity1}, which makes it possible to selectively include a subset of nodes that are likely to belong to the same community
as contexts of a current node. Sections \ref{ss:merw} and \ref{ss:ca2nv-objective} explain these two approaches.

%combines the proper community in-formations with the learned node2vec representations for better contextualization of a node in a network which in-turn lead to better identifications of community structures in a network. In addition, the entropy based biased random walk in the process of node2vec also affects the improvement of the performances for the right identification of actual community structure present in a network. The following sections provides the details of the entropy based random walk and its association to the objective function of node2vec for better contextualization of a node in a network.

\subsection{Maximal-Entropy Random Walk (MERW)} \label{ss:merw}

In contrast to the uniform walk of Equation \ref{eq:urw}, maximal-entropy random walk (MERW) is characterized by a stochastic matrix that maximises entropy of a set of paths (node sequences) with a given length and end-points \cite{Ochab2013}, leading to the stochastic matrix
\begin{equation}
P_{uv}=\frac{A_{uv}}{\lambda}\frac{\psi_v}{\psi_u} \label{eq:merw},
\end{equation}
where $\lambda$ denotes the largest eigenvalue of the adjacency matrix $A$, with $\psi_v$ and $\psi_u$ the
$v^{th}$ and $u^{th}$ components of the corresponding eigenvectors.
Applying Frobenius-Perron theorem proves that the probability of visiting a node $u_n$ after $n$ time steps starting from node $u_1$ depends only on the number of steps and the two ending points. It is, however, independent of the intermediate nodes \cite{Parry1964}, i.e.,
\begin{equation}
P(u_1,\ldots u_n)=\prod_{i=1}^{n-1}P_{u_i,u_{i+1}}=\frac{1}{\lambda^n}\frac{\psi_{u_1}}{\psi_{u_n}}.  
\end{equation}
Thus, the next node to visit in MERW relies on
uniformly selecting the node from alternative paths of a given length and end-points.
The study in \cite{Centrality2011} shows that the stationary distribution attained by MERW better preserves centrality than URW, thus resulting in random walks that tend to be more local as shown in \cite{MERW-Loc}. In the context of our problem, MERW based random walk initiated from a node of a community is more likely to remain within the confinements of the same community, as compared to URW.

We already mentioned that node2vec uses a different approach of DFS/BFS based walk to construct the set of contexts for a node. We hypothesize that replacing the DFS/BFS based walk based neighborhood with MERW potentially results in a lower likelihood of spanning across communities from a peripheral node of a community. Specifically, with respect to Equation \ref{eq:node2vec}, if nodes $u$ and $v$ belong to different (true) communities, likelihood of
including the node $v$ in the neighborhood of $u$, $\mathcal{N}(u)$, is potentially low. This results in a low likelihood of including the term $P(y=1|\vec{u},\vec{v})$ in the objective of Equation \ref{eq:node2vec}, i.e. associating nodes across two different communities as a positive example for training node representations.

%\section{Modularity-aware Node Embedding}
%TODO

\subsection{Modified Objective Function for Node Embedding} \label{ss:ca2nv-objective}

In this section, we describe a two-step approach to node embedding that is likely to preserve the community structure of the discrete space of an input graph in the output embedded space.
The first step involves applying a combinatorial community detection algorithm that operates on the discrete input space to obtain an optimal partition, as per the objective function of the combinatorial approach, e.g. modularity \cite{clauset2004finding} or InfoMap \cite{rosvall2008maps}. Formally,
\begin{equation}
\mathcal{C}: G=(V,E) \mapsto \{V_i\}_{i=1}^p,\,\mathrm{s.t.}\, \cup_{i=1}^p V_i = V \label{eq:combpart}, 
\end{equation}
i.e., a combinatorial algorithm partitions the vertex set, $V$, of a graph into $p$ distinct communities.

In the second step, for obtaining the node embedding instead of providing as input the unpartitioned graph (as in standard approaches), we rather input the partitioned set of vertices obtained from Equation \ref{eq:combpart}.
Based on the supplied partition, we modify the objective function of node2vec (Equation \ref{eq:node2vec}) to address differently the two types of positive node association within a context, i.e., one, where
node pairs belong to the same community (partition) as induced by the partition,
and the other, where they belong to different communities.
We put more emphasis on the first case than on the second one. Formally speaking,
%
\begin{comment}
\begin{equation}
\begin{split}
J(\theta|\mathcal{C}) & = \alpha
\sum_{u \in V_i} \sum_{v \in \mathcal{N}(u) \cap V_i}P(y=1|\vec{u},\vec{v}) + \\
& (1-\alpha)\sum_{u \in V_i} \sum_{v \in \mathcal{N}(u)-V_i}\!\!\!\!\!\!\!\!\!P(y=1|\vec{u},\vec{v}) +
\sum_{u} \sum_{\bar{v} \in \bar{\mathcal{N}}(u)}\!\!\!\!P(y=0|\vec{u},\vec{\bar{v}}) \label{eq:mnode2vec},
\end{split}
\end{equation}
\end{comment}
%
\begin{equation}
J(\theta|\mathcal{C}) = \alpha
\sum_{u \in V_i} \sum_{v \in \mathcal{N}(u) \cap V_i}\!\!\!\!\!\!\!\!P(y=1|\vec{u},\vec{v}) +
(1-\alpha)\sum_{u \in V_i} \sum_{v \in \mathcal{N}(u)-V_i}\!\!\!\!\!\!\!\!\!P(y=1|\vec{u},\vec{v}) +
\sum_{u} \sum_{\bar{v} \in \bar{\mathcal{N}}(u)}\!\!\!\!P(y=0|\vec{u},\vec{\bar{v}}) \label{eq:mnode2vec},
\end{equation}
where
the first component indicates those cases where $u$ and $v$ are predicted to be a part of the same community by a combinatorial algorithm $\mathcal{C}$,
the second component indicates the ones where
$u$ and $v$ are predicted to be a part of different communities as per $\mathcal{C}$, and
$\alpha \in [0,1]$ indicates a relative importance of the first component over the second (specifically for our experiments, we set $\alpha=0.8$).

\input{plot_of_main_ideas/plot_main_idea_points.tex} 

The intuition behind Equation \ref{eq:mnode2vec} is to rely on two different sources of information, for determining the similarities between node pairs. The risk of only using the random walk based information is that a random walk initiated from the periphery of a community is likely to visit a peripheral node of a different community. Considering these cases as positive examples in the node2vec objective could result in falsely embedding two such nodes close to each other, in which case, it would be difficult for a downstream clustering algorithm, such as K-means, to assign them into two distinct clusters.
However, using the additional information about the estimated communities is likely to identify these false cases and hence down-weight them in the embedding objective. Note that the contribution to the objective for node pairs belonging to different communities is still positive (i.e. $y=1$) as compared to the negative samples ($y=0$) when a vertex is selected at random from outside the set of visited nodes.

Returning to the earlier example graph of node2vec (i.e. Figure \ref{fig:rwalk-n2v}), Figure \ref{fig:rwalk-can2v} illustrates the idea of differently treating node pairs based on the induced node partition information.
The figure shows that since the vertex pair (u8, v7) is a part of two different estimated communities, its detrimental effect as a false positive example (in terms of the ground-truth) on the embedding objective is reduced by weighing its contribution down to $1-\alpha$ as per Equation \ref{eq:mnode2vec}.

\subsection{Maximum Entropy Random Walk for Node Embedding}

From Figure \ref{fig:rwalk-can2v}, we observe that modified node2vec objective of Equation \ref{eq:mnode2vec} makes the same mistake as node2vec for those node pairs where the estimated community partition do not align with the true community information, e.g. for the (u7, v8) case where the initial partition predicts that they belong to the same community (whereas as per the ground-truth, they belong to two different ones).

To further improve the robustness of estimating the node vectors, we propose to use to incorporate the MERW based neighborhood construction within Equation \ref{eq:mnode2vec}. Since MERW is
likely to preserve centrality with respect to a community
\cite{MERW-Loc}, a walk is likely to be confined within the same community.
Specifically, to incorporate the maximum-entropy objective, instead of using the original graph weights for obtaining the return and in-out parameter based random walk of node2vec \cite{Grover:2016}, we apply spectral analysis on the original graph (Equation \ref{eq:merw}) to modify edge weights before applying the $p$-$q$ biased walk.

\para{Variants of our proposed approaches}
In addition to taking as input the number of clusters, $K$, our proposed approach of modifying the node2vec objective (which we denote as \textbf{\CAnv} or community aware node2vec) also takes as input the partition induced by a combinatorial method, leading to a likely different output partitioning. Consequently, we report results on three different instances of \CAnv~(one each for CNM, Louvian and LPA).
In a similar manner, we report results with the
three different cases (each corresponding
to a combinatorial community detection approach) for the MERW-based node2vec (denoted as \textbf{\Mnv}) and community-aware MERW based node2vec (combination of both MERW based context construction and community partition driven modified node2vec objective), which we denote as \textbf{\MCAnv}.

\section{Experiment Setup}  \label{sec:experimentsetup}

In this section, we describe the setup of our experiments for community detection. Specifically, we describe the graphs (both real-life and synthetically generated) used for our experiments, the methods investigated and the evaluation measures undertaken.

\subsection{Datasets}  \label{ss:datasets}

We conduct experiments on a range of different undirected and unweighted networks of varying sizes (number of nodes) and densities (relative number of edges with respect to a complete graph). All the graphs that we experimented with are associated with the ground-truth community information.

\input{tables/dataset}

\para{Real-life Networks}

First, we perform experiments on three relatively small-scale standard benchmark networks for community detection. The first among these is the `karate club' \footnote{\url{https://networkdata.ics.uci.edu/data.php?id=105}} graph, which comprises 34 nodes and 78 edges, where every node represents a member of a karate club at an American university. If two members are observed to have social interactions within or away from the karate club, they are connected by an edge.
Another small network that we experiment with is the
`dolphin network' comprising 62 nodes that represent bottlenose dolphins living in Doubtful Sound, New Zealand. The edges in this graph (159 in total) represent associations between dolphin pairs that were observed to be more frequent than the occasional expectation. The third network used in our experiments is the network of American football games between Division IA colleges during regular season of Fall 2000 \cite{girvan2002community}.

In addition to these two small networks, we also conduct experiments on a large network \cite{snapnets}, namely the
%Amazon product purchase network and the
Youtube user group network and the DBLP network \cite{yang2015defining}. These networks are undirected and unweighted and they are selected from different application domains. The overview of these networks are presented Table 1.
%
%We have also tested the experiments on three real world networks viz. Amazon and DBLP \cite{yang2015defining, harenberg2014community,snapnets}. These networks are undirected and unweighted and they are selected from different application domains. The overview of these networks are presented Table 1.
%
%In the Amazon product purchase network, nodes represent products and an edge exists between two products if they are frequently purchased together. Each product (i.e. node) belongs to one or more product categories. Each ground-truth community is defined using hierarchically nested product categories that share a common function \cite{yang2015defining}. 
%
In the DBLP network, a bibliographic network of computer science publications, a node represents an author, and an edge between two nodes represents co-authorship. Ground-truth communities are defined as sets of authors who has published at least once in the same venue denoting a common topical interest \cite{mislove-2007-socialnetworks}.
Each user in the Youtube network is considered to be a node and the friendship between two users is denoted as edge. Moreover, an user can create a closed group by inviting his friends. Such groups are considered as ground-truth communities \cite{mislove-2007-socialnetworks}. 

%\cite{yang2015defining} quantified the quality (termed as \emph{goodness}) of the community structure of a network as a function of how compact are the communities and how well are they connected internally while being relatively well-separated from the rest of the network.
The study in \cite{harenberg2014community} observed that the
community structure of a real-life entire network (e.g. the Youtube network) is approximated by about top $5000$ communities, following which
%average goodness metric of the top $k$ communities starts monotonically decreasing for values of $k$ higher than $5000$. The study \citep{yang2015defining} also shows that for the sake of comparing community detection approaches, it suffices to restrict the computation to the most representative communities of a large graph (typically the top $5000$ communities).
%Following this observation, we in our experiments,
we also restrict our experiments to the top $5000$ communities for the
Youtube network.
%
%Therefore they have implemented some community detection algorithms using different goodness metrics on the top 5000 communities of some of the networks described above. Eventually, they obtained nice results in terms of finding communities in those networks. Following the same idea we have used only the top 5000 ground-truth communities of each of these networks in the experimental evaluation.
%
%The networks described above have several connected components and each connected component consisting of more than 3 nodes are considered as a separate ground-truth community.
%

\para{Synthetic Networks} 

In addition to the real-life networks, we also
conduct experiments on synthetic networks, generated with
the standard LFR (Lancichinetti-Fortunato-Radicchi) mechanism
%to generate graphs with good community structures
%(i.e. relatively dense communities with sparse inter-community links)
\cite{lancichinetti2008benchmark}.
An important parameter in the power law based LFR generative mechanism is
%The LFR model involve with a set of parameters which controls the network topology. In this model, degree distribution and community size distribution follow power laws with exponents $\gamma$ and $\beta$, respectively. Furthermore, we can also specify the other parameters such as number of vertices $n$, average degree $k_{avg}$, maximum degree $k_{max}$, minimum community size $c_{min}$, maximum community size $c_{max}$, and mixing parameter $\mu$. We vary these parameters depending on our experimental needs. The critical parameter is
the mixing parameter $\mu$, which indicates the proportion of relationships a node shares with other communities. We have used here $\mu=0.3$ for the quantitative analysis of the variants of our proposed approaches.
%As prescribed by \cite{lancichinetti2008benchmark}, we used $\mu=0.1$.
%Six artificial networks are produced for experimental evaluations using the following parameter setting, $\gamma = -2$,$\beta = -1$ $\mu = 0.01$ as mentioned by Lancichinetti et al. \cite{lancichinetti2008benchmark}.
%Table \ref{tab:description_real_world_data_sets} provides the details of the LFR parameters to generate the artificial networks for our experiments.
To reduce randomization effects of the artificially generated networks, we report the average results (over a set of $100$ instances) obtained with each competing method.
%The results presented on the synthetic networks are the average of 100 runs to reduce the effect of random assumptions. 
%
Table \ref{tab:description_real_world_data_sets} summarizes the different networks used in our experiments.

\subsection{Methods Investigated}
The objective of our experiments is to investigate if our proposed node embedding approaches (with the MERW and the modified objective function based on a combinatorial approach) is able to outperform standard embedding and combinatorial approaches for community detection.
As our combinatorial baselines to community detection, we employ
two methods that use the modularity score to greedily aggregate nodes into communities, and a random walk based method.
Specifically, as the modularity score based approaches, we use
CNM algorithm \cite{Newman2004modularity1}, which operates on a graph as a whole, and the Louvian algorithm \cite{blondel2008fast} (denoted as `LV' in our experiments), which successively coarsens a graph for community aggregation.
As the random-walk based baseline, we employ the LPA and the INFOMAP algorithms (abbreviated as IMap). As the final combinatorial approach, we employ SCDA \cite{prat2014high}, which uses triangle analysis to detect communities.
It is to be noted that these combinatorial baseline approaches automatically estimate the optimal number of clusters (communities) by making use of a global heuristic function representing the quality of the community structure.

As a node embedding based baseline for community detection, we employ a two-step method, the first step applying node2vec to obtain the embedded node representations of a graph, followed by conducting K-means on the node vectors to predict the communities (each cluster corresponding to a community). We denote this baseline as \textbf{n2v} in our experiments. In contrast to the combinatorial approaches, for K-means clustering, the number of communities needs to be provided as input.
For each combinatorial community detection algorithm, as mentioned before, we employ the number of communities obtained by each as the value of $K$ in the clustering based approach.

As the first community-aware baseline,
we employ the community-aware node embedding approach (\textbf{COM-E}) proposed in \cite{Cavallari:2017}, which jointly maximizes the node-context similarities \cite{Grover:2016} along with a node vector's similarity with its cluster centroid. Since a parameter to COM-E is the number of clusters ($K$), which is not known a-priori, we tested COM-E with values of $K$ obtained with each combinatorial partition, e.g. CNM etc.

\input{tables/karate.tex}

\subsection{Evaluation Measures}

In this section, we describe the evaluation metrics used to measure the community detection effectiveness. The networks that we experimented with are associated with
ground-truth, i.e., for each node it is known in which community (or communities) it belongs to. Since the task and its evaluation is analogous to evaluating clustering effectiveness
using the ground-truth cluster labels (in our case, a community is analogous to a cluster), we make use of the standard cluster evaluation metrics to evaluate community detection. Such standard clustering metrics can broadly be categorized into two different categories, namely the ones which are based on the correctness of the pairwise cluster assignments, and
the ones that are based on how (truly) homogeneous are the constructed clusters.  

\input{tables/dolphin.tex}

Among pairwise decision based metrics, we use the Omega-Index ($\Omega$) \cite{collins1988omega}, and the mean F-score \cite{yang2013overlapping}. Among the homogeneity based ones, we use `Normalized Mutual Information' (\NMIMAX), and its square-root variant, \NMISQRT ~\cite{strehl2002cluster}. We now briefly describe each metric.
%Previous research These metrics are common among researchers used by a number of researchers \cite{yang2012community} to measure the equality of different communities produced by an algorithm using the ground-truth communities of the network.

\input{tables/football.tex}

\para{Omega Index}
The Omega-Index \cite{collins1988omega} (reported as $\Omega$ in our experiments) is a generalization of the `Adjusted Rand Index' (ARI) metric \cite{hubert1985comparing} applicable to overlapping communities). It is based on counting the number of pairs of elements occurring in exactly the same number of clusters as in the number of categories and adjusted to the expected number of such pairs. 

\input{tables/youtube.tex}

\para{Mean F-Score}
The mean F-score (reported as F1 in our experiments) is a commonly used metric to measure clustering effectiveness. It is a combination of the precision and recall of the correctenss of the pairwise node assignments to communities, i.e., a combination of how many of the pairs predicted to belong to the same community are true (precision), vs. how many of these true pairs are actually detected out of the total number of known ones (recall) \cite{yang2013overlapping}.

\input{tables/dblp.tex}

\para{Normalized Mutual Information (NMI)}
Mutual Information (MI) is evaluated by
aggregating the overlap (in terms of the number of common elements) between a predicted partition, $P$ and a ground-truth cluster $C$ \cite{strehl2002cluster}. Formally, given a set of clusters
$\mathcal{C}=\cup \{C\}$ and a set of estimated partitions $\mathcal{P}=\cup \{P\}$,
\[
I(\mathcal{C},\mathcal{P})=\sum_{C \in \mathcal{C}}\sum_{P \in \mathcal{P}} |C \cap P| \log \frac{|C \cap P|}{|C| |P|},
\]
For easier interpretation and comparisons, the mutual information value computed this way is usually normalized in the
range of $[0,1]$ \cite{nmi}. The two most common ways of this normalization involve using as normalization constants either the average or the maximum values
of the respective entropy measures for the true and the predicted partition sets, i.e. $\mathcal{C}$ and $\mathcal{P}$. In our experiments, we denote these
two measures as $\sqrt{\mathrm{NMI}}$, and NMI respectively.
Since NMI is a measure of the similarity between the true and estimated clusters, a high value of NMI corresponds to a better clustering effectiveness.

\input{tables/lfr1k_new.tex}

\input{fig-defs/lfr1k.tex}

\section{Results}   \label{sec:results}
We start this section by reporting the best results obtained by each community detection approach on the real-world and synthetic networks, described in Section \ref{ss:datasets}. We then investigate the effect of varying the parameters on the node embedding approaches.

%%%%%%%%%%%%%%%%%%%%%%%%%%%%%%%%%%%%%%
\subsection{Comparisons between node embedding based community detection methods}

We now present the quantitative comparisons between the effectiveness of the different approaches investigated. Tables \ref{tab:karateres}, \ref{tab:dolphinres} and \ref{tab:footballres} show the community detection results obtained on the three small networks, namely, Karate-club, Dolphin and Football networks. Tables \ref{tab:youtuberes} and \ref{tab:dblpres} show the results corresponding to the two large large networks viz. Youtube and DBLP networks respectively. Table \ref{tab:lfr1kres} shows the community detection results on the synthetic network LFR-1K network.

%as mentioned in Table \ref{tab:description_real_world_data_sets}.

In addition to showing the results for each combinatorial approach, we make use of each partition as inputs to the variants of our proposed approach \CAnv~and \MCAnv~(yielding the different rows with different values of $K$). We also show the results yielded with the oracle settings (i.e. the number of desired partitions being set to the number of ground-truth communities in the embedding based approaches), which gives an estimate of the upper bound of the community detection effectiveness. 

For the baseline COM-E \cite{Cavallari:2017}, we conducted a grid search with all the distinct values of $K$ obtained from the combinatorial methods. Note that since the COM-E baseline does not directly depend on the partition itself but rather only on the desired number of clusters, Table \ref{tab:karateres} shows only $3$ results for these baselines (each corresponding to a unique value of $K$ in `Comb' group).

Also note that the oracle settings for n2v baseline and variants of our approaches (\Mnv, \CAnv and \MCAnv) yield different results for different partitions induced by different methods even if the number of clusters is identical (this happens because the results depend on the partition set itself and not just on the value of $K$). On the other hand COM-E yields a single set of results corresponding to one value of $K$ (the optimal one). 

The oracle cases in Tables \ref{tab:karateres}-\ref{tab:lfr1kres} correspond to the variants of our proposed approach only, i.e., where we substitute the true community information within the node2vec objective of Equation \ref{eq:mnode2vec} to yield an upper bound in terms of effectiveness. The rest of the cells are left empty (colored gray), e.g., the cells corresponding to a combinatorial heuristic or node2vec. 

%whereas Tables \ref{tab:results_lfr500_all} and \ref{tab:results_lfr1000_all} report results for the LFR-based synthetic networks.
%
In general, the following trends can be observed from the results.
First, K-means on embedded node vectors mostly outperforms purely combinatorial approaches (e.g. CNM, LPA, etc.), more so for the large network.
%(e.g. the increase of \NMIMAX~from $0.5056$ with LPA to $0.7549$ with n2v).
Second, we observe that the use of MERW for node embedding mostly improves community detection effectiveness,
%e.g. \NMIMAX~ of MERW-n2v ($0.9308$) is higher than that of n2v ($0.9248$).
which confirms our hypothesis that maximum entropy based random walk is likely to include nodes of the same community in the contexts that are used to train the embedding.

Third, we observe that incorporating the partition information within the objective of node embedding results (i.e. \CAnv) substantially improves the results in comparison to n2v and \Mnv~(both do not use the partition information), as seen from the presence of most bold-faced and underlined values in \CAnv~and \MCAnv~groups through Tables \ref{tab:karateres}-\ref{tab:lfr1kres}. This suggests that the initial partitions information's has an positive effect to improve the performances over node embedding results. From Tables \ref{tab:karateres}-\ref{tab:lfr1kres}, it is clear that the LPA and IMap partitions information's within the node embedding results mostly outperform the other partitioning results such as CNM, LV and SCDA. Same conclusions can be done over the oracle settings as shown in \ref{tab:karateres}-\ref{tab:lfr1kres}.

Finally, a combination of both partition awareness and MERW for constructing node contexts for training node vectors is seen to improve community detection effectiveness further (\MCAnv~ results corresponding to the Tables \ref{tab:karateres}-\ref{tab:lfr1kres}). The results also show that our proposed methods outperform COM-E, the reason for which could be attributed to the fact that addressing similarities of a node with its cluster centroid results in a `smoothing' effect due to the averaging. On the other hand, our approach is more fine grained since we model the similarities between node pairs
(Equation \ref{eq:mnode2vec}).

%\subsection{Graphical Analysis}

%\subsection{Parameter Sensitivity Analysis}

\input{fig-defs/mu_2.tex}

\input{fig-defs/mu_4.tex}
\input{fig-defs/mu_6.tex}
\input{fig-defs/mu_8.tex}

\subsection{Parameter Sensitivity with respect to Graph Size and Community Structure}

In this section, we investigate how effective are our proposed community detection approaches over a wide range of graphs of different sizes and community structures in comparison to the baseline approach, n2v.
To simulate graphs with different inherent community structural properties, we generate a range of graphs with different sizes and community structures (e.g. different interlink patterns and densities of communities).

As networks of different sizes, we investigate with 1000 and 5000 nodes, respectively denoted as LFR-1K and LFR-5K. As possible choices of the mixing parameter, $\mu$, we use $\mu=\{0.2, 0.4, 0.6, 0.8\}$. Recall from the LFR generation methodology that the value of $\mu$ indicates what fraction of edges of every node is
connected with nodes in other communities, the remaining ($1-\mu$) being connected with nodes of the same community. Consequently, a lower value of $\mu \in [0,1]$ results in more dense and distinguishable communities \cite{lancichinetti2008benchmark}.

Figures \ref{fig:sensitivitylfr1000} and \ref{fig:sensitivitylfr5000} show the comparisons between n2v, \Mnv, \CAnv~and \MCAnv~in terms of the metrics - NMI, $\Omega$-index and F-scores for a range of LFR networks with varying node sizes and range of community structures.
As the node partitioning heuristic for the experiments with the community aware embedding approaches - \CAnv~ and \MCAnv, we specifically used the IMAP partitioning heuristic.

As a general trend, it can be observed that the community effectiveness decreases with increasing values of $\mu$, i.e., as expected when the communities are more distinguishable, node embedding approaches tend to perform better. Moreover, it is also seen that over a range of different networks (with varying community structure), \MCAnv~and \CAnv~perform better than n2v and \Mnv, which indicates that community aware embedding approaches perform consistently better than the community agnostic ones. 
 
%The plots in Figure \ref{fig:sensitivitylfr1000} and \ref{fig:sensitivitylfr5000} show that it is clear that the performance of the proposed variant (viz. \CAnv~ and \MCAnv) always better than the performance of n2v and \Mnv~ for different values of the mixing parameter $\mu$ over the network. As the value of $\mu$ increases the measure of performance of all the competitive methods sharply decreases. The difference between the measure of performances of the competitive methods become insignificant for higher values of $\mu$.

\subsection{Sensitivity with respect to Embedding Parameters}

We now investigate how do the embedding parameters,
namely $p$ (return parameter), $q$ (in-out parameter) and $ws$ (context size), affect the community detection effectiveness of one of our proposed variants (specifically, \MCAnv~ which in most cases yielded the best community detection effectiveness in the results reported in Tables \ref{tab:karateres}-\ref{tab:lfr1kres}.
Specifically, each embedding walk parameter $p$ and $q$ was set to a low ($0.1$) and a high value ($0.5$) thus resulting in $4$ different combinations, e.g. a low value of $p$ and a high value of $q$ and so on. Additionally, to see how does relative changes to the embedded vectors (induced by the parameters) behave with respect to different community structure and sizes, we vary the mixing parameter $\mu$ on the LFR-1K network.

General observations from Figures \ref{fig:sensitivity02}-\ref{fig:sensitivity08} are as follows. First, it is evident that small and large window sizes (respectively, $5$ and $100$ lead to the best results).
While on the one hand, large window sizes may result in arranging the vectors in an embedded space into a small number of large clusters, small window sizes, on the other hand, is likely to result in a large number of small clusters. Since the true number of clusters is neither very large nor very small, this gives the algorithm a better chance to either merge the small groups of embedded node vectors into a larger cluster (of medium size) or decompose a larger one into a smaller one (again of medium size).  
The cases, when the embedded vectors are neither too far apart nor too close to each other (mid-range window sizes) turn out to be difficult cases for K-means on the embedded space.

Second, the community detection effectiveness decreases with increasing values of $\mu$, as can be seen from the progressively decreasing values of the evaluation metrics from Figure \ref{fig:sensitivity02} to \ref{fig:sensitivity04} and so on. This shows that with increasing values of the LFR generation parameter, $\mu$, the communities become more indistinguishable from each other (they in fact do not behave like true communities any more), as a result of which, it turns out progressively difficult for the node embedding algorithm to arrange them into well-separable clusters in the embedded space thus decreasing the effectiveness of community detection.

Third, it can be seen that a combination of high values for both $p$ (the return parameter) and $q$ (the in-out parameter) turns out to be the best for community detection. This indicates a conflicting setting for the parameters, because while on the one hand a high value of $p$ intends to explore yet unseen nodes of a graph more (thus likely spreading away the walk to large distances from the initial node), a high value of $q$ in is likely to restrain the walk to the local vicinity of its starting node. Since we apply a maximum-entropy based random walk (Equation \ref{eq:merw}), it turns out that it helps in \emph{exploring} the graph in an ambitious (high value of $p$) yet a controlled manner (through the centrality heuristic of MERW).

A low value of $p$ coupled with a high value of $q$ turns out to be the best for the case $\mu=0.8$. Since in these simulated graphs, there are a higher number of edges spanning across the communities, the embedding algorithm needs to be more restrictive in its exploration phase to avoid likely visits to nodes of other communities.

\section{Conclusions and Future work}   \label{sec:conclusions}
In this paper, we proposed a novel community detection algorithm, which relies on leveraging information from an estimated partition of the network (into communities) for the purpose of embedding the nodes of the network.
This is likely to alleviate the problem of a random-walk based context construction for node embedding, as the random walk may eventually lead to including nodes from different communities in the context of a node.
We hypothesize that our proposed community-aware embedding algorithm leads to better separability of the
embedded node vectors, which in turn increases the clustering (community detection) effectiveness on this embedded space.
%We also employ a maximum-entropy based random walk for constructing the context of a node, which is likely to better preserve the centrality of the communities of a network in the embedded space.
%
%combinatorial approaches into the objective function for learning node representations. 
Further, we investigated a maximal entropy based random walk (which is known to preserve locality), and its combination with the partition augmented embedding objective.

The results of our experiments on a number of real-life and synthetic networks demonstrate that - i) including the combinatorial heuristics-based partitional information helps improve community detection effectiveness, ii) the global perspective introduced by the maximum entropy based random walk helps makes it more likely to confine the walk within communities thus leading to improved embedding, and iii) a combination of the above two works well in practice to yield more effective node vectors.

As future work, we plan to investigate how node embedding in combination with combinatorial approaches for graph partition could be used to detect communities in a dynamically evolving network, i.e., a network whose structure can evolve with time, e.g. a social-media friendship network or a citation network. We also plan to incorporate other sources of information, such as the attributes associated with the nodes of a network (e.g. the text description of a node), into the combinatorial heuristics or as a part of the maximum-entropy random walk objective so as to eventually benefit community detection in such node attributed networks.

%% file: plot_of_main_ideas/plot_main_idea.tex
\begin{figure}[t]
 \centering
\begin{subfigure}[t]{.6\textwidth}
    \centering
    \includegraphics[width=\columnwidth]{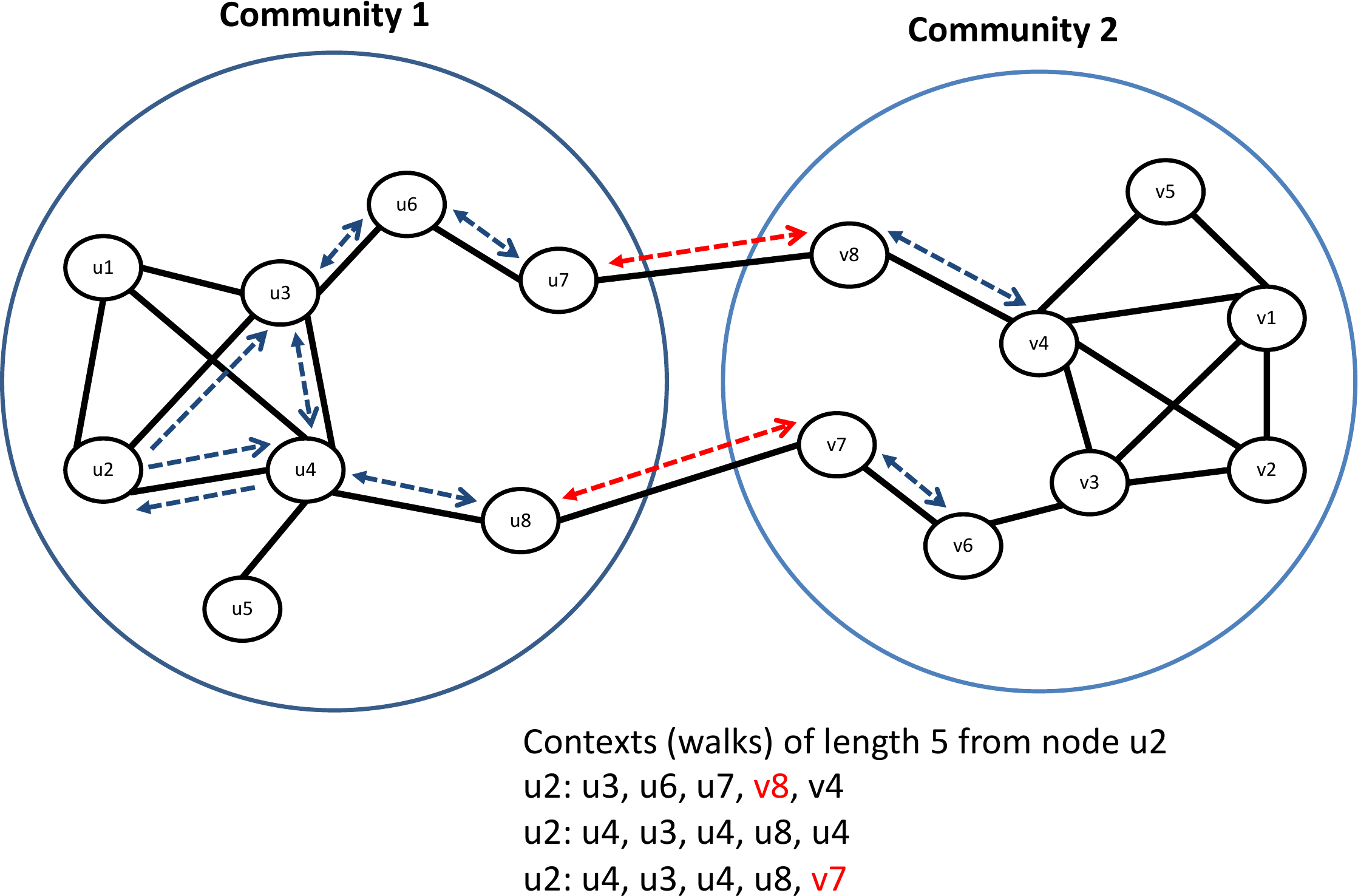}
    \caption{Random walks in an example network with 2 true communities. \label{fig:rwalk-n2v}}
    ~
    \end{subfigure}
\begin{subfigure}[t]{.37\textwidth}
    \centering
    \includegraphics[width=\columnwidth]{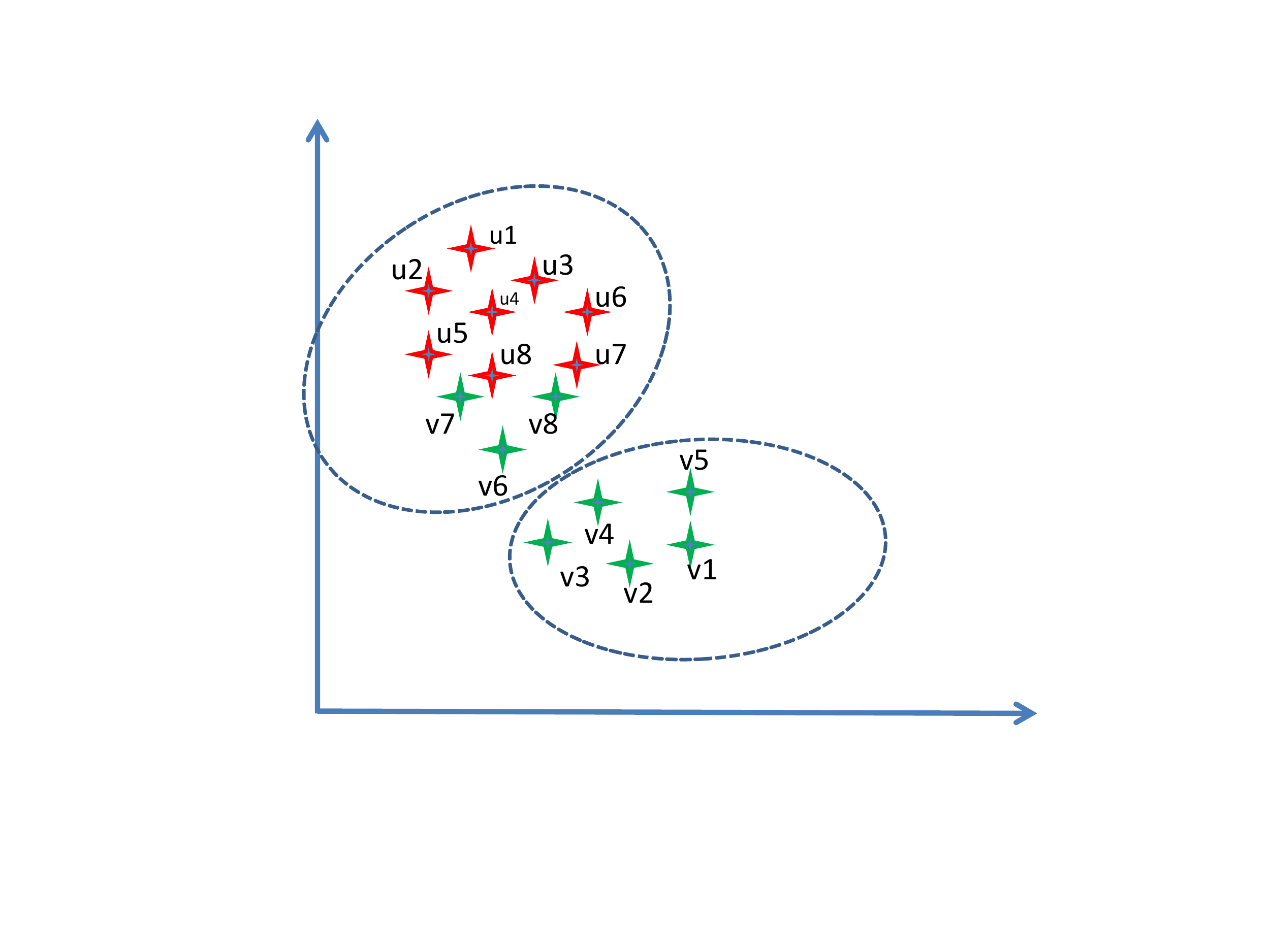}
    \caption{Communities detected with K-means on embedded space.}
    \end{subfigure}
% \begin{subfigure}[t]{.32\textwidth}
%     \centering
%     \includegraphics[width=\columnwidth]{sensitivity-figs/lfr_mu_2_1000_mean_fscore.pdf}
%     \caption{Mean F-score }
%     \end{subfigure}
\caption{Schematic illustration of the limitations of node2vec embedding.
The left figure shows that node2vec random walk can span across nodes of two different (ground-truth) communities (shown with red arrows). Random walks constructed this way may lead to low distances between nodes of two different communities, as shown by the distance between u8 and v7 in the figure on the right.
As a result, K-means clustering on these vectors could lead to non-homogeneous clusters.
\label{fig:n2v-schematic}
}

\end{figure}

%% file: plot_of_main_ideas/plot_main_idea_points.tex
\begin{figure}[t]
 \centering
\begin{subfigure}[t]{.6\textwidth}
    \centering
    \includegraphics[width=\columnwidth]{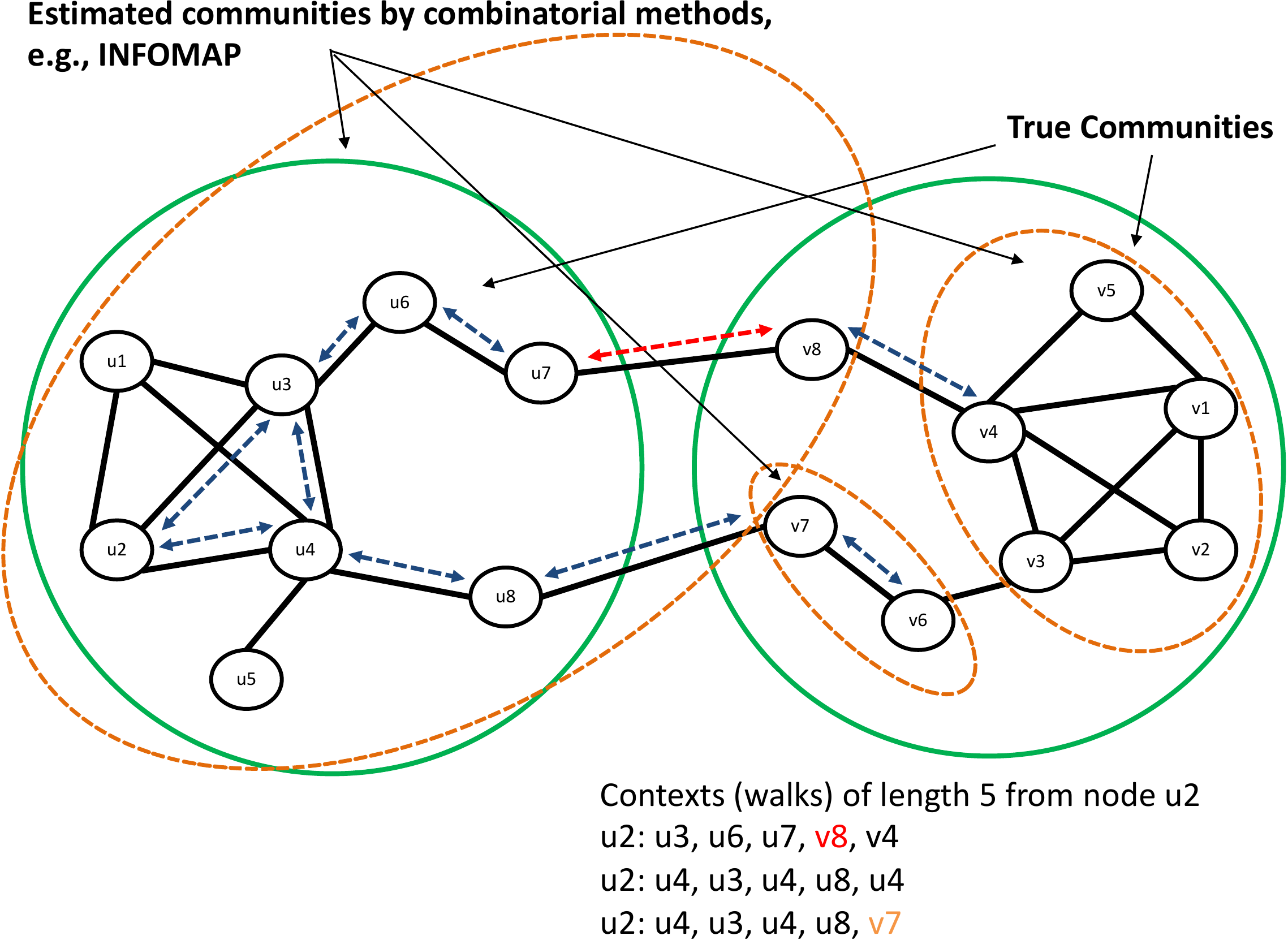}
    \caption{Example graph showing 2 true (green circles) and 3 communities (amber dotted ovals) estimated with a combinatorial heuristic, e.g., modularity \cite{Newman2004modularity1}).
    To denote a different contributing factor from the node pair (u8, v7), the node v7 is shown in amber. The node v8 is shown in red to denote that it is a false positive (similar to the node2vec case of Figure \ref{fig:rwalk-n2v}). 
    \label{fig:rwalk-can2v}
    }
    \end{subfigure}
    ~
\begin{subfigure}[t]{.37\textwidth}
    \centering
    \includegraphics[width=\columnwidth]{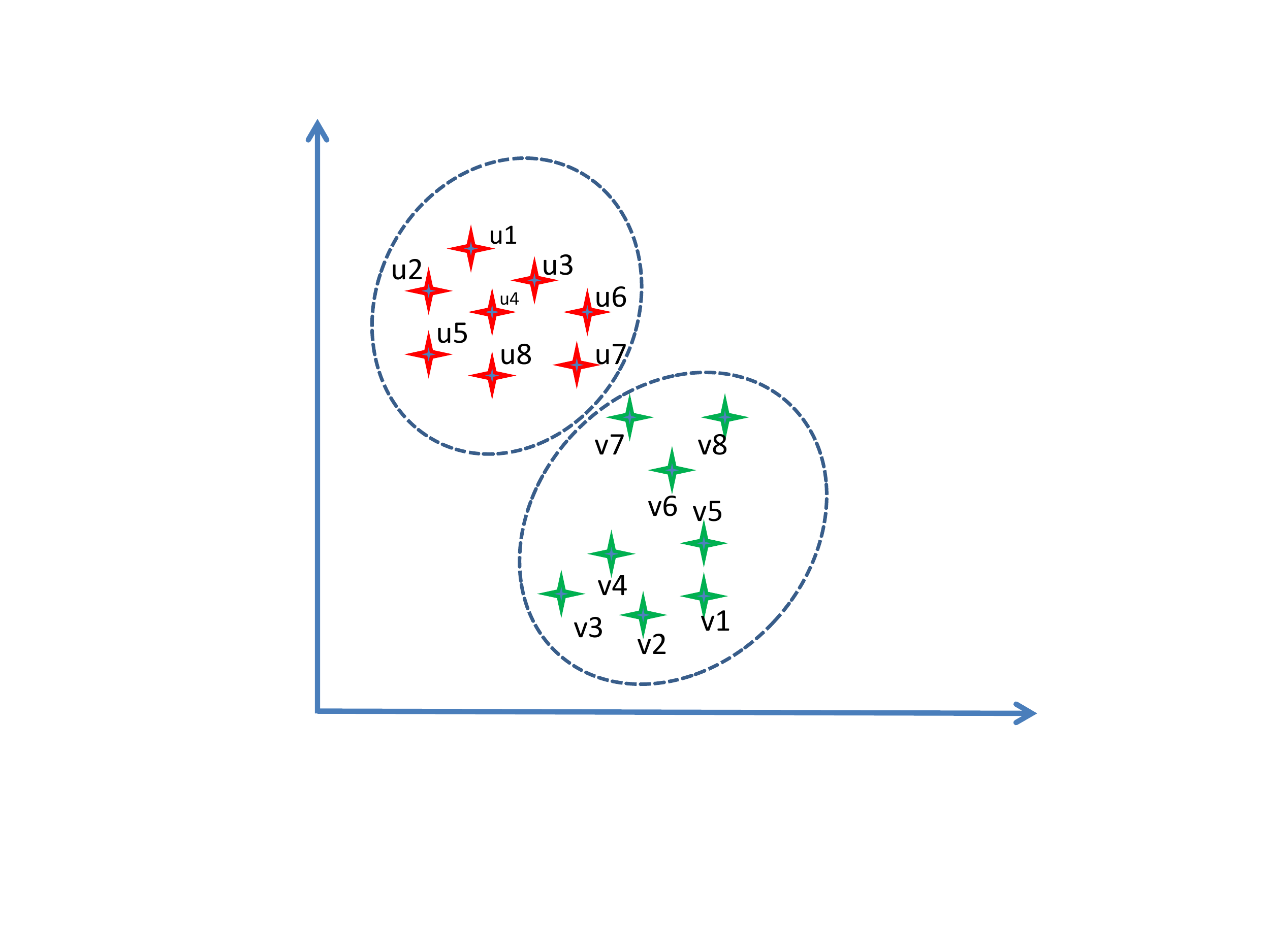}
    \caption{Communities detected with K-means on embedded space.}
    \end{subfigure}
\caption{
Schematic illustration of our proposed community-aware node embedding.
The left figure shows that using the estimated partition, the modified algorithm treats the node pairs (u7, v8) and (u8, v7) differently. Since the latter (a true negative example) is a part of two different estimated communities, its detrimental effect as a positive example on the embedding objective is reduced by weighing its contribution down to $1-\alpha$ as per Equation \ref{eq:mnode2vec}.
%The algorithm still makes the same mistake as node2vec for a node pair where the estimated community partition does not align with the true community information, e.g. for the (u7, v8) case where the initial partition predicts that they belong to the same community.
As a result, the embedded space is likely to mostly preserve the topological structure of the original network, and K-means clustering on these vectors are likely to lead to homogeneous clusters.
\label{fig:can2v-schematic}
}
\end{figure}

%% file: tables/dataset.tex
\begin{table}[t]
\centering
%\setlength{\tabcolsep}{4pt}
%\begin{tabular}{|c|c|c|c|c|c|c|c|c|c|}
\scriptsize
%\begin{tabular}{@{}l@{~~}c@{~~~~}c@{~~~~}c@{~~~}r@{~~}r@{~~~}r@{~~~}r@{~~~}r@{~~~}r@{}}
\begin{tabular}{lcccrrrrrr}
\toprule
Network & $|V|$ & $|E|$ & $\rho$ & $C_{num}$ & $C_{max}$ & $C_{min}$ & $k_{max}$ & $k_{avg}$ & $C_{avg}$ \\
\midrule
Karate  & 34 & 78 & 0.2288 & 2 & 18 & 16 & 17 & 4.588 & 17.00 \\
Dolphin  & 62 & 159 & 0.1278 & 2 & 42 & 20 & 12 & 5.129 & 31.00 \\
Football  & 115 & 613 & 0.1101 & 12 & 13 & 5 & 12 & 10.66 & 9.58 \\
%Amazon  & 16716 & 48739 & 0.0126 & 5000 & 328 & 3 & 51 & 5.831 & 13.49 \\
Youtube  & 39481 & 224235 & 0.0036 & 5000 & 2217 & 2 & 1575 & 11.26 & 14.59 \\
DBLP  & 93432 & 335520 & 0.0011 & 5000 & 7556 & 6 & 213 & 7.182 & 22.45 \\
%LFR500 & 500 & 1410 & 0.1408 & 38 & 45 & 2 & 49 & 5.64 & 13.16 \\
LFR1000 & 1000 & 3973 & 0.0651 & 40 & 97 & 5 & 50 & 7.946 & 25.01 \\
\bottomrule
\end{tabular}
\caption{Overview of a number of benchmark real-life networks used in our experiments. Acronyms: 
$\rho$ (Minimum Internal Density),
$C_{num}$ (\#communities),
$C_{max}$ (Maximum Community Size),
$C_{min}$ (Minimum Community Size),
$k_{max}$ (Maximum Degree),
$k_{avg}$ (Average Degree),
$C_{avg}$ (Average Community Size).\\
\label{tab:description_real_world_data_sets}
}
\end{table}

%% file: tables/karate.tex
\begin{table}[t]
\scriptsize
\begin{center}
\caption{Comparison of different community detection methods on the Karate Club network. For each method, we conduct a grid search over possible partitions induced by a set of combinatorial approaches (shown separated with horizontal lines).
The results of our proposed approaches are separated from the baselines with a double line. 
The best results for each group are bold-faced. In addition, the best results with a metric (maximum over an entire column) are underlined.
We follow the same convention through Tables \ref{tab:dolphinres}-\ref{tab:lfr1kres}.
\label{tab:karateres}
}
%\setlength{\tabcolsep}{2pt}
%{
%\begin{tabular}{|c|c|c|c|c|c|c|}
%\begin{tabular}{@{}l@{~~}l@{~~}l@{~~}c@{~~}c@{~~}c@{~~}c@{~~}c@{~~} c@{~~}c@{~~}c@{~~}c@{}}
\begin{tabular}{lllccccccccc}
\toprule
\multicolumn{2}{c}{Method} & & \multicolumn{4}{c}{Results} & \multicolumn{4}{c}{Oracle Results} \\
%\cmidrule(r){1-2}
\cmidrule(r){4-7} \cmidrule(r){8-11}
\multicolumn{2}{c}{} & $K$ & \NMIMAX & \NMISQRT & $\Omega$ & F1 & \NMIMAX & \NMISQRT & $\Omega$ & F1\\
\toprule
\multirow{5}{*}{\Combinatorial} &	CNM &	4 & 0.4518 &	0.6231 &	0.4909 &	0.7518 &
\nores \\
 &	LV	& 4	& 0.4426	& 0.6100	& 0.4619	& 0.7507 & \nores  \\
 &	LPA &	3 & \textbf{0.5902} &	0.7058	& \textbf{0.7022}	& \textbf{0.8677} & \nores \\
 &	IMap &	3 & 0.5890 &	\textbf{0.7072} &	\textbf{0.7022} & \textbf{0.8677}  & \nores \\
 &	SCDA &	8	& 0.4255	& 0.6523	& 0.4828 &	0.7405 & \nores \\
\midrule
\multirow{5}{*}{n2v}	& CNM &	4 &	0.4541 & 0.5943 &	0.4876 &	0.7582 & \nores \\
	& LV	& 4	& 0.4541 &	0.5943 &	0.4876 &	0.7582 & \nores \\
	& LPA &	3 &	\textbf{0.5224} & \textbf{0.6175}	& \textbf{0.5836}	& \textbf{0.8251} &
\textbf{0.6325} & \textbf{0.6417} &	\textbf{0.6877} &	\textbf{0.9170} \\
	& IMap &	3 &	\textbf{0.5224} & \textbf{0.6175} &	\textbf{0.5836} &	\textbf{0.8251} &\nores\\
	& SCDA &	8	& 0.3526 &	0.5600 &	0.3241 &	0.6305 &\nores \\
\midrule
\multirow{3}{*}{COM-E} & &	3 &	\textbf{0.5308} &	\textbf{0.6160} &	\textbf{0.5987} & \textbf{0.8394} & \nores \\
& & 4 & 0.4757 &	0.6134 &	0.5014 &	0.7622 & \textbf{0.6846} & \textbf{0.6925} & {0.7259} & \textbf{0.9282} \\
% \multirow{3}{*}{\textbf{0.6846}} &
% \multirow{3}{*}{\textbf{0.6925}} &
% \multirow{3}{*}{\textbf{0.7259}} &
% \multirow{3}{*}{\textbf{0.9282}} \\
% &	& 4 &	0.4757 &	0.6134 &	0.5014 &	0.7622 & & & & \\
 &	& 8	& 0.3668 &	0.5637 &	0.3562 &	0.6493 & \nores \\
\midrule
\midrule
\multirow{5}{*}{\CAnv} &	CNM &	4	& 0.5489 &	0.6816 &	0.5974 &	0.8076 &
\textbf{\underline{0.8051}} & \textbf{\underline{0.8074}} &	\textbf{\underline{0.8546}} &	\textbf{\underline{0.9637}} \\
 &	LV &	4 &	0.5419 &	0.6982 &	0.5989 &	0.8058 &
0.7503 &	0.7536 &	0.8071 &	0.9506 \\
 &	LPA &	3	& \textbf{\underline{0.6629}} &	\textbf{\underline{0.7506}} &	\textbf{\underline{0.7514}} &	\textbf{\underline{0.8957}} &
0.7845 &	0.7858 &	0.8491 &	0.9622 \\
 &	IMap &	3 &	0.6176	& 0.7011	& 0.7101	& 0.8770 &
0.7892 &	0.7917 &	0.8435 &	0.9601 \\
 &	SCDA &	8 &	0.4523 &	0.6638 &	0.4808 &	0.7340 &
0.7711 &	0.7754 &	0.8217 &	0.9551 \\
\midrule
\multirow{5}{*}{\Mnv} &	CNM &	4 & 0.4672 &	0.6085 &	0.5104 &	0.7244 &\nores\\
 &	LV &	4	& 0.4672 &	0.6085 &	0.5104 &	0.7244 &\nores	 \\
 &	LPA &	3 	& \textbf{0.5573} &	\textbf{0.6531} &	\textbf{0.6316} &	\textbf{0.8394} &
\textbf{0.6911} &	\textbf{0.6983} & \textbf{0.7463} &	\textbf{0.9328} \\
 &	IMap &	3	& \textbf{0.5573} &	\textbf{0.6531} &	\textbf{0.6316} &	\textbf{0.8394} &\nores\\
 &	SCDA &	8 &	0.3633 &	0.5675 &	0.3538 &	0.6532 &\nores\\
\midrule
\multirow{5}{*}{\MCAnv} & CNM &	4 &	0.5212 &	0.6614 &	0.6016 &	0.8148 &
0.7398 &	0.7451 &	0.7990 &	0.9485 \\
 &	LV &	4 &	0.5590 &	0.7160 &	0.6292 &	0.8186 &
0.7208 &	0.7256 &	0.7739 &	0.9427 \\
 &	LPA &	3 &	\textbf{0.6457} &	\textbf{0.7332} &	\textbf{0.7344} &	\textbf{0.8904} &
\textbf{0.7631} &	\textbf{0.7662} &	\textbf{0.8270} &	\textbf{0.9564}\\
 &	IMap &	3	& 0.6164 &	0.7096 &	0.7141 &	0.8789 &
0.7556 &	0.7577 &	0.8141 &	0.9531 \\
 &	SCDA	& 8	& 0.4700	& 0.6674	& 0.5181	& 0.7492 &
0.7563 &	0.7623 &	0.8065 &	0.9509 \\
\bottomrule
\end{tabular}
%}
\end{center}
\end{table}

%% file: tables/dolphin.tex
\begin{table}[t]
\scriptsize
\begin{center}
\caption{Comparison of different community detection methods on Dolphin network. \label{tab:dolphinres}}
%\setlength{\tabcolsep}{2pt}
%{
%\begin{tabular}{|c|c|c|c|c|c|c|}
%\begin{tabular}{@{}l@{~~}l@{~~}l@{~~}c@{~~}c@{~~}c@{~~}c@{~~}c@{~~} c@{~~}c@{~~}c@{~~}c@{}}
\begin{tabular}{lllccccccccc}
\toprule
\multicolumn{2}{c}{Method} & & \multicolumn{4}{c}{Results} & \multicolumn{4}{c}{Oracle Results} \\
%\cmidrule(r){1-2}
\cmidrule(r){4-7} \cmidrule(r){8-11}
\multicolumn{2}{c}{} & $K$ & \NMIMAX & \NMISQRT & $\Omega$ & F1 & \NMIMAX & \NMISQRT & $\Omega$ & F1\\
\toprule
\multirow{5}{*}{\Combinatorial} &	CNM &	4 &	0.4225 &	0.5867 &	0.4509 &	0.7860 & \nores \\
 &	LV	&	6 &	0.3201 &	0.5312 &	0.2709 &	0.6334  & \nores \\
 &	LPA&	4 &	\textbf{0.4960} &	\textbf{0.6712} &	\textbf{0.5090} &	\textbf{0.8031}  & \nores \\
 &	IMap &	6 &	0.3932 &	0.6210 &	0.3605	& 0.7015   & \nores \\
 &	SCDA &	26 &	0.2173 &	0.4564	& 0.1137	& 0.4463 	 &  \nores \\
\midrule
\multirow{5}{*}{n2v}	& CNM &	4 &	\textbf{0.4362} &	\textbf{0.6149} &	\textbf{0.4469} &	\textbf{0.7517}	 & \nores \\
	& LV& 6 &	0.3568 &	0.5449 &	0.3436 &	0.6798	 &  \nores	 \\
	& LPA &	4 &	\textbf{0.4362} &	\textbf{0.6149} &	\textbf{0.4469} &	\textbf{0.7517}	 & \textbf{0.7875} &	\textbf{0.8019} &	\textbf{0.8600} &	\textbf{0.9652} \\
	& IMap &	6 &	0.3568 &	0.5449 &	0.3436 &	0.6798	 &  \nores	\\
	& SCDA &	26 &	0.2322 & 0.4780 &	0.1156 &	0.4576	 &  \nores	\\
\midrule
\multirow{3}{*}{COM-E} & &	4 &	\textbf{0.4351}	& \textbf{0.6157}	& \textbf{0.4679}	& \textbf{0.7628}	 & \nores \\
%\multirow{3}{*}{\textbf{0.8074}} &
%\multirow{3}{*}{\textbf{0.8188}} &
%\multirow{3}{*}{\textbf{0.8789}} &
%\multirow{3}{*}{\textbf{0.9698}} \\
 &	& 6 & 0.3273	& 0.5014	& 0.3115	& 0.6857	 & \textbf{0.8074}& \textbf{0.8188} & \textbf{0.8789}& \textbf{0.9698} \\
 &	& 26	& 0.2329	& 0.4752	& 0.1183	& 0.4633 	 & \nores \\
\midrule
\multirow{5}{*}{\CAnv} &	CNM &	4 &	0.4652 &	0.6378 &	0.4945 &	0.7869	 &  0.8274 &	0.8391 &	0.8909 & 0.9730	 \\
 &	LV &	6 &	0.3659 &	0.5796	& 0.3396 &	0.6747	 &  0.8312 &	0.8427 &	0.8972 &	0.9746	 \\
 &	LPA &	4 &	\textbf{0.5016} &	\textbf{\underline{0.6731}} &	\textbf{\underline{0.5460}} &	\textbf{\underline{0.8263}}	 & \textbf{\underline{0.8563}} &	\textbf{\underline{0.8659}} &	\textbf{\underline{0.9160}} &	\textbf{\underline{0.9793}}	 \\
 &	IMap &	6 &	0.4090 &	0.6229 &	0.3898 &	0.7220 &  0.8333 &	0.8446 &	0.8974 &	0.9746	 \\
 &	SCDA  &	26 &	0.2471 &	0.4924 &	0.1372 &	0.4812	 &  0.8351 &	0.8462 &	0.8974 &	0.9746	 \\
\midrule
\multirow{5}{*}{\Mnv} &	CNM &	4 &	\textbf{0.4455} &	\textbf{0.6239} &	\textbf{0.4567} &	\textbf{0.7693}	 &  \nores	 \\
 &	LV  &	6 &	0.3816 &	0.5815 &	0.3764 &	0.7086	 & \nores	 \\ 
 &	LPA & 4	& \textbf{0.4455}	& \textbf{0.6239}	& \textbf{0.4567}	& \textbf{0.7693}	 &
 \textbf{0.8087} &	\textbf{0.8202}	& \textbf{0.8788} &	\textbf{0.9699}	 \\ 
 &	IMap &	6	& 0.3816	& 0.5815	& 0.3764	& 0.7086	 & \nores		 \\
 &	SCDA & 26 &	0.2380	& 0.4845	& 0.1201	& 0.4581	 & \nores		 \\
\midrule
\multirow{5}{*}{\MCAnv} & CNM & 4	& 0.4570	& 0.6181	& 0.4811	& 0.7837	 & 0.8235 &	0.8354 &	0.8909 & 0.9730	 \\
 &	LV &	6 &	0.3734	& 0.5791	& 0.3550	& 0.6974	 &
0.8284 &	0.8399 &	0.8914 &	0.9730	 \\
 &	LPA &	4	& \textbf{\underline{0.5025}}	& \textbf{0.6720}	& \textbf{0.5342}	& \textbf{0.8101}	 &
 \textbf{0.8344} &	\textbf{0.8457} &	0.8972 &	\textbf{0.9746}	 \\
 &	IMap &	6	& 0.4064	& 0.6223 &	0.3954 &	0.7117	 &
  0.8299 &	0.8399 &	0.8971 &	0.9745	 \\
 &	SCDA	&	26 &	0.2472 &	0.4955 &	0.1302 &	0.4731	 &
0.8341 &	\textbf{0.8457} &	\textbf{0.8974} &	\textbf{0.9746}	 \\
\bottomrule
\end{tabular}
%}
\end{center}
\end{table}

%% file: tables/football.tex
\begin{table}[t]
\scriptsize
\begin{center}
\caption{Comparison of different community detection methods on Football network. \label{tab:footballres}}
%\setlength{\tabcolsep}{2pt}
%{
%\begin{tabular}{|c|c|c|c|c|c|c|}
%\begin{tabular}{@{}l@{~~}l@{~~}l@{~~}c@{~~}c@{~~}c@{~~}c@{~~}c@{~~} c@{~~}c@{~~}c@{~~}c@{}}
\begin{tabular}{lllccccccccc}
\toprule
\multicolumn{2}{c}{Method} & & \multicolumn{4}{c}{Results} & \multicolumn{4}{c}{Oracle Results} \\
%\cmidrule(r){1-2}
\cmidrule(r){4-7} \cmidrule(r){8-11}
\multicolumn{2}{c}{} & $K$ & \NMIMAX & \NMISQRT & $\Omega$ & F1 & \NMIMAX & \NMISQRT & $\Omega$ & F1\\
\toprule
\multirow{5}{*}{\Combinatorial} &	CNM &	6 &	0.5906 &	0.7022 &	0.4741 &	0.6907	 & \nores \\
 &	LV	&	10 &	0.8560 &	0.8885 &	0.8069 &	0.8920  & \nores \\
 &	LPA&	11 &	0.8815 &	0.9013 &	0.8465 &	0.9123  & \nores \\
 &	IMap &	12 &	\textbf{0.9187} &	\textbf{0.9193} &	\textbf{0.8967}	& \textbf{0.9212}   & \nores \\
 &	SCDA &	15 &	0.8894 &	0.9142	& 0.8484	& 0.8961 	 & \nores \\
\midrule
\multirow{5}{*}{n2v}	& CNM &	6 &	0.5830 &	0.6924 &	0.4557 &	0.6740	 & \nores \\
	& LV&	10 &	0.7892 &	0.8262 &	0.6965 &	0.8294	 & \nores	 \\
	& LPA &	11 &	0.8192 &	0.8411 &	0.7349 &	0.8433	 & \textbf{0.8444} &	\textbf{0.8505} &	\textbf{0.7588} &	\textbf{0.8494} \\
	& IMap &	12 &	\textbf{0.8406} &	0.8516 &	0.7534 &	\textbf{0.8480}	 & \nores	\\
	& SCDA &	15 &	0.8366 & \textbf{0.8566} &	\textbf{0.7602} &	0.8378	 & \nores	\\
\midrule
\multirow{5}{*}{COM-E} & &	6 &	0.5837	& 0.6958	& 0.4576	& 0.6735	 & \nores \\
% \multirow{5}{*}{\textbf{0.8485}} &
% \multirow{5}{*}{\textbf{0.8577}} &
% \multirow{5}{*}{\textbf{0.7674}} &
% \multirow{5}{*}{\textbf{0.8529}} \\
 &	& 10 & 0.7899	& 0.8280	& 0.6999	& 0.8335	 & \nores \\
 &	& 11	& 0.8233	& 0.8448	& 0.7387	& 0.8489 	 & \textbf{0.8485} & \textbf{0.8577} & \textbf{0.7674} & \textbf{0.8529} \\
 &	& 12 & \textbf{0.8443}	& \textbf{0.8576}	& \textbf{0.7609}	& \textbf{0.8519}	 & \nores \\
&	& 15 &	0.8364	& 0.8470	& 0.7504	& 0.8365	 & \nores \\
\midrule
\multirow{5}{*}{\CAnv} &	CNM &	6 &	0.6175 &	0.7331 &	0.5029 &	0.6995	 &  0.8730 &	0.8854 &	0.8109 & 0.8821	 \\
 &	LV &	10 &	0.8614 &	0.8850	& 0.7964 &	0.8917	 &  0.8792 &	0.8875 &	0.8233 &	0.8914	 \\
 &	LPA &	11 &	0.8881 &	0.8895 &	0.8381 &	0.8920	 &  0.8849 &	0.8908 &	0.8269 &	0.8939	 \\
 &	IMap &	12 &	\textbf{\underline{0.9208}} &	\textbf{\underline{0.9198}} &	\textbf{\underline{0.8995}} &	\textbf{\underline{0.9229}}	 &  \textbf{\underline{0.9008}} &	\textbf{\underline{0.9083}} &	\textbf{\underline{0.8596}} &	\textbf{\underline{0.9019}}	 \\
 &	SCDA  &	15 &	0.8857 &	0.8893 &	0.8591 &	0.8947	 &  0.8856 &	0.8925 &	0.8428 &	0.8951	 \\
\midrule
\multirow{5}{*}{\Mnv} &	CNM &	6 &	0.6003 &	0.7147 &	0.4844 &	0.6895	 &  \nores \\
 &	LV  &	10 &	0.8026 &	0.8405 &	0.7263 &	0.8426	 & \nores		 \\ 
 &	LPA & 11	& 0.8395	& 0.8582	& 0.7772	& 0.8666	 &
 \textbf{0.8553} &	\textbf{0.8619}	& \textbf{0.7828} &	\textbf{0.8679}	 \\ 
 &	IMap &	12	& \textbf{0.8551}	& 0.8614	& \textbf{0.7834}	& \textbf{0.8691}	 & \nores		 \\
 &	SCDA & 15 &	0.8463	& \textbf{0.8669}	& 0.7809	& 0.8491	 & \nores		 \\
\midrule
\multirow{5}{*}{\MCAnv} & CNM & 6	& 0.6241	& 0.7417	& 0.5068	& 0.6992	 & 0.8829 &	0.8898 &	0.8296 & 0.8951	 \\
 &	LV &	10 &	0.8753	& 0.8997	& 0.7998	& 0.8972	 &
0.8751 &	0.8848 &	0.8194 &	0.8887	 \\
 &	LPA &	11	& 0.8905	& 0.9006	& 0.8487	& 0.9128	 &
 0.8916 &	0.8985 &	0.8395 &	0.8996	 \\
 &	IMap &	12	& \textbf{0.9022}	& \textbf{0.9056} &	\textbf{0.8897} &	\textbf{0.9139}	 &
  \textbf{0.8972} &	\textbf{0.9016} &	\textbf{0.8497} &	\textbf{0.9001}	 \\
 &	SCDA	&	15 &	0.8909 &	0.8985 &	0.8494 &	0.8972	 &
 0.8911 &	0.8966 &	0.8452 &	0.8964	 \\
\bottomrule
\end{tabular}
%}
\end{center}
\end{table}

%% file: tables/youtube.tex
\begin{table}[t]
\scriptsize
\begin{center}
\caption{Comparison of different community detection methods on Youtube network. \label{tab:youtuberes}}
%\setlength{\tabcolsep}{2pt}
%{
%\begin{tabular}{|c|c|c|c|c|c|c|}
%\begin{tabular}{@{}l@{~~}l@{~~}l@{~~}c@{~~}c@{~~}c@{~~}c@{~~}c@{~~} c@{~~}c@{~~}c@{~~}c@{}}
\begin{tabular}{lllccccccccc}
\toprule
\multicolumn{2}{c}{Method} & & \multicolumn{4}{c}{Results} & \multicolumn{4}{c}{Oracle Results} \\
%\cmidrule(r){1-2}
\cmidrule(r){4-7} \cmidrule(r){8-11}
\multicolumn{2}{c}{} & $K$ & \NMIMAX & \NMISQRT & $\Omega$ & F1 & \NMIMAX & \NMISQRT & $\Omega$ & F1\\
\toprule
\multirow{5}{*}{\Combinatorial} &	CNM &	1271 &	0.3680 &	0.5817 &	0.0666 &	0.2773	 &
\nores \\
 &	LV	&	890 &	0.3875 &	0.5898 &	0.0848 &	0.2809  & \nores \\
 &	LPA&	2695 &	0.5065 &	0.6770 &	0.0893 &	0.4109  & \nores \\
 &	IMap &	2954 &	\textbf{0.7450} &	\textbf{0.8001} &	\textbf{0.1874}	& \textbf{0.4396}   & \nores \\
 &	SCDA &	25435 &	0.7173 &	0.8190	& 0.0262	& 0.3334 	 & \nores \\
\midrule
\multirow{5}{*}{n2v}	& CNM &	1271 &	0.7295 &	0.7726 &	0.1713 &	\textbf{0.4174}	 &  \nores \\
	& LV&	890 &	0.7037 &	0.7597 &	\textbf{0.1799} &	0.4036	 &  \nores	 \\
	& LPA &	2695 &	0.7549 &	0.7717 &	0.1172 &	0.3954	 & \textbf{0.7975} &	\textbf{0.8098} &	\textbf{0.1062} &	\textbf{0.4432} \\
	& IMap &	2954 &	\textbf{0.7626} &	0.7747 &	0.1246 &	0.4015	 &  \nores	\\
	& SCDA &	25435 &	0.7337 & \textbf{0.8284} &	0.0173 &	0.3429	 &  \nores	\\
\midrule
\multirow{5}{*}{COM-E} & &	1271 &	0.7312	& 0.7739	& 0.1459	& 0.4105	 & \nores \\
% \multirow{5}{*}{\textbf{0.7965}} &
% \multirow{5}{*}{\textbf{0.8113}} &
% \multirow{5}{*}{\textbf{0.1109}} &
% \multirow{5}{*}{\textbf{0.4521}} \\
 &	& 890 & 0.7078	& 0.7632	& \textbf{0.1697}	& 0.4075	 & \nores \\
 &	& 2695	& 0.7874	& 0.7957	& 0.1157	& \textbf{0.4306} 	 & \textbf{0.7965} & \textbf{0.8113} & \textbf{0.1109} & \textbf{0.4521} \\
 &	& 2954 & \textbf{0.7881}	& 0.7892	& 0.1242	& 0.4182	 & \nores \\
&	& 25435&	0.7382	& \textbf{0.8348}	& 0.0201	& 0.3605	 & \nores \\
\midrule
\multirow{5}{*}{\CAnv} &	CNM &	1271 &	0.7361 &	0.7760 &	0.1665 &	0.4210	 &  0.8003 &	0.8131 &	0.1143 & 0.4511	 \\
 &	LV &	890 &	0.7104 &	0.7621	& \textbf{0.1828} &	0.4061	 &  0.8013 &	0.8138 &	\textbf{\underline{0.1229}} &	0.4537	 \\
 &	LPA &	2695 &	0.7887 &	0.7903 &	0.1214 &	0.4316	 &  \textbf{\underline{0.8023}} &	0.8140 &	0.1183 &	0.4524	 \\
 &	IMap &	2954 &	\textbf{0.7908} &	0.7950 &	0.1253 &	\textbf{0.4391}	 &  0.7984 &	0.8137 &	0.1165 &	\textbf{\underline{0.4543}}	 \\
 &	SCDA  &	25435 &	0.7417 &	\textbf{\underline{0.8359}} &	0.0221 &	0.3682	 &  0.7982 &	\textbf{0.8143} &	0.1125 &	0.4531	 \\
\midrule
\multirow{5}{*}{\Mnv} &	CNM &	1271 &	0.7307 &	0.7730 &	0.1685 &	0.4167	 &  \nores		 \\
 &	LV  &	890 &	0.7034 &	0.7592 &	\textbf{0.1769} &	0.4018	 & \nores	 \\ 
 &	LPA & 2695	& 0.7819	& 0.7933	& 0.1341	& 0.4317	 &
 \textbf{0.7979} &	\textbf{0.8102}	& 0.1094 &	\textbf{0.4448}	 \\ 
 &	IMap &	2954	& \textbf{0.7900}	& 0.7974	& 0.1276	& \textbf{0.4366}	 &
\nores	 \\
 &	SCDA & 25435 &	0.7399	& \textbf{0.8353}	& 0.0200	& 0.3601	 &
\nores		 \\
\midrule
\multirow{5}{*}{\MCAnv} & CNM & 1271	& 0.7365	& 0.7754	& 0.1574	& 0.4196	 & 0.7989 &	0.8132 &	0.1143 & 0.4510	 \\
 &	LV &	890 &	0.7104	& 0.7641	& \textbf{\underline{0.1882}}	& 0.4075	 &
\textbf{0.7994} &	0.8134 &	0.1151 &	0.4527	 \\
 &	LPA &	2695	& 0.7874	& 0.7960	& 0.1285	& 0.4346	 &
 0.7991 &	0.8116 &	\textbf{0.1206} &	0.4531	 \\
 &	IMap &	2954	& \textbf{\underline{0.7955}}	& 0.7997 &	0.1275 &	\textbf{\underline{0.4399}}	 &
  0.7957 &	0.8114 &	0.1019 &	0.4482	 \\
 &	SCDA	&	25435 &	0.7415 &	\textbf{0.8357} &	0.0210 &	0.3642	 &
 \textbf{0.7994} &	\textbf{\underline{0.8148}} &	0.1149 &	\textbf{0.4537}	 \\
\bottomrule
\end{tabular}
%}
\end{center}
\end{table}

%% file: tables/dblp.tex
\begin{table}[t]
\scriptsize
\begin{center}
\caption{Comparison of different community detection methods on DBLP network. \label{tab:dblpres}}
%\setlength{\tabcolsep}{2pt}
%{
%\begin{tabular}{|c|c|c|c|c|c|c|}
%\begin{tabular}{@{}l@{~~}l@{~~}l@{~~}c@{~~}c@{~~}c@{~~}c@{~~}c@{~~} c@{~~}c@{~~}c@{~~}c@{}}
\begin{tabular}{lllccccccccc}
\toprule
\multicolumn{2}{c}{Method} & & \multicolumn{4}{c}{Results} & \multicolumn{4}{c}{Oracle Results} \\
%\cmidrule(r){1-2}
\cmidrule(r){4-7} \cmidrule(r){8-11}
\multicolumn{2}{c}{} & $K$ & \NMIMAX & \NMISQRT & $\Omega$ & F1 & \NMIMAX & \NMISQRT & $\Omega$ & F1\\
\toprule
\multirow{5}{*}{\Combinatorial} &	CNM &	1515 &	0.5087 &	0.6026 &	0.9764 &	0.3220	 & \nores
 \\
 &	LV	&	594 &	0.5466 &	0.6144 &	0.9659 &	0.3122  & \nores \\
 &	LPA&	9239 &	0.6426 &	0.7637 &	\textbf{\underline{0.9767}} &	0.4385  & \nores \\
 &	IMap &	6577 &	\textbf{0.6536} &	\textbf{0.7723} &	\textbf{\underline{0.9767}}	& \textbf{0.4436}   & \nores \\
 &	SCDA &	28253 &	0.6115 &	0.7719	& 0.9766	& 0.3792 	 & \nores \\
\midrule
\multirow{5}{*}{n2v}	& CNM &	1515 &	0.6529 &	0.7032 &	0.9764 &	0.3341	 &  \nores  \\
	& LV&	594 &	0.6451 &	0.6572 &	0.9757 &	0.3073	 &  \nores \\
	& LPA &	9239 &	0.6451 &	0.7618 &	\textbf{\underline{0.9767}} &	0.4442	 & \textbf{0.6504} &	\textbf{0.7477} &	\textbf{0.9766} &	\textbf{0.3954} \\
	& IMap &	6577 &	\textbf{0.6570} &\textbf{0.7741} &	0.9766 &	0.4437	 &  \nores	\\
	& SCDA &	28253 &	0.6281 & 0.7714 &	0.9761 &	0.4007	 &  \nores	\\
\midrule
\multirow{5}{*}{COM-E} & &	1515 &	0.6533	& 0.7036	& 0.9764	& 0.3349	 & \nores \\
% \multirow{5}{*}{\textbf{0.6413}} &
% \multirow{5}{*}{\textbf{0.7411}} &
% \multirow{5}{*}{\textbf{0.9765}} &
% \multirow{5}{*}{\textbf{0.3879}} \\
 &	& 594 & 0.6450	& 0.6576	& 0.9752	& 0.3089	 & \nores \\
 &	& 9239	& 0.6399	& 0.7638	& \textbf{\underline{0.9767}}	& \textbf{0.4449} 	 & \textbf{0.6413} & \textbf{0.7411} & \textbf{0.9765} & \textbf{0.3879} \\
 &	& 6577 & \textbf{0.6556}	& \textbf{0.7755}	& 0.9766	& 0.4429	 & \nores \\
&	& 28253 &	0.6264	& 0.7711	& 0.9764	& 0.4000	 & \nores \\
\midrule
\multirow{5}{*}{\CAnv} &	CNM &	1515 &	0.6553 &	0.7040 &	0.9764 &	0.3369	 &  0.6513 &	0.7487 &	0.9765 & 0.3996	 \\
 &	LV &	594 &	0.6521 &	0.6559	& 0.9758 &	0.3126	 &  0.6489 &	0.7413 &	0.9763 &	0.3894	 \\
 &	LPA &	9239 &	0.6492 &	0.7687 &	\textbf{\underline{0.9767}} &	\textbf{0.4489}	 &  0.6510 &	0.7438 &	0.9763 &	0.3983	 \\
 &	IMap &	6577 &	\textbf{0.6580} &	\textbf{\underline{0.7758}} &	\textbf{\underline{0.9767}} &	0.4466	 &  \textbf{0.6600} &	\textbf{\underline{0.7499}} &	\textbf{\underline{0.9767}} &	\textbf{\underline{0.4018}}	 \\
 &	SCDA  &	28253 &	0.6295 &	0.7725 &	0.9762 &	0.4016	 &  0.6456 &	0.7395 &	0.9762 &	0.3782	 \\
\midrule
\multirow{5}{*}{\Mnv} &	CNM &	1515 &	0.6534 &	0.7036 &	0.9764 &	0.3350	 &  \nores	 \\
 &	LV  &	594 &	0.6455 &	0.6585 &	0.9752 &	0.3093	 & \nores		 \\ 
 &	LPA & 9239	& 0.6437	& 0.7635	& \textbf{\underline{0.9767}}	& 0.4399	 &
 \textbf{0.6507} &	\textbf{0.7477}	& \textbf{0.9765} &	\textbf{0.3967}	 \\ 
 &	IMap &	6577	& \textbf{0.6560}	& \textbf{0.7746}	& 0.9766	& \textbf{0.4438}	 &
 \nores		 \\
 &	SCDA & 28253 &	0.6279	& 0.7713	& \textbf{\underline{0.9767}}	& 0.4001	 &
  \nores		 \\
\midrule
\multirow{5}{*}{\MCAnv} & CNM & 1515	& 0.6556	& 0.7048	& 0.9764	& 0.3357	 & 0.6492 &	0.7479 &	\textbf{0.9766} & 0.3981	 \\
 &	LV &	594 &	0.6539	& 0.6601	& 0.9758	& 0.3136	 &
0.6464 &	0.7429 &	0.9764 &	0.3885	 \\
 &	LPA &	9239	& 0.6472	& 0.7668	& \textbf{\underline{0.9767}}	& \textbf{\underline{0.4498}}	 &
 0.6508 &	0.7465 &	0.9764 &	0.3979	 \\
 &	IMap &	6577	& \textbf{\underline{0.6582}}	& \textbf{\underline{0.7758}} &	\textbf{\underline{0.9767}} &	0.4486	 &
  \textbf{\underline{0.6611}} &\textbf{	0.7488} &	0.9764 &	\textbf{0.3998}	 \\
 &	SCDA	&	28253 &	0.6299 &	0.7747 &	0.9764 &	0.4018	 &
 0.6445 &	0.7418 &	0.9765 &	0.3809	 \\
\bottomrule
\end{tabular}
%}
\end{center}
\end{table}

%% file: tables/lfr1k_new.tex
\begin{table}[t]
\scriptsize
\begin{center}
\caption{Comparison of different community detection methods on LFR-1K network. \label{tab:lfr1kres}}
%\setlength{\tabcolsep}{2pt}
%{
%\begin{tabular}{|c|c|c|c|c|c|c|}
%\begin{tabular}{@{}l@{~~}l@{~~}l@{~~}c@{~~}c@{~~}c@{~~}c@{~~}c@{~~} c@{~~}c@{~~}c@{~~}c@{}}
\begin{tabular}{lllccccccccc}
\toprule
\multicolumn{2}{c}{Method} & & \multicolumn{4}{c}{Results} & \multicolumn{4}{c}{Oracle Results} \\
%\cmidrule(r){1-2}
\cmidrule(r){4-7} \cmidrule(r){8-11}
\multicolumn{2}{c}{} & $K$ & \NMIMAX & \NMISQRT & $\Omega$ & F1 & \NMIMAX & \NMISQRT & $\Omega$ & F1\\
\toprule
\multirow{5}{*}{\Combinatorial} &	CNM &	16 &	0.6137 &	0.7325 &	0.4123 &	0.6249	 &
\nores \\
 &	LV	&	25 &	0.8532 &	0.9133 &	0.8752 &	0.8861  & \nores \\
 &	LPA&	38 &	\textbf{0.9314} &	\textbf{0.9430} &	\textbf{0.9110} &	\textbf{0.9243}  & \nores \\
 &	IMap &	48 &	0.9275 &	0.9418 &	0.9101	& 0.9229  & \nores \\
 &	SCDA &	466 &	0.6196 &	0.7857	& 0.2935	& 0.5335 	 & \nores \\
\midrule
\multirow{5}{*}{n2v}	& CNM &	16 &	0.7032 &	0.7837 &	0.5944 &	0.7108	 &  \nores  \\
	& LV&	25 &	0.8372 &	0.8709 &	0.7825 &	0.8475	 &   \nores	 \\
	& LPA &	38 &	\textbf{0.8748} &	\textbf{0.8832} &	\textbf{0.8285} &	\textbf{0.8847}	 & \textbf{0.8757} &	\textbf{0.8878} &	\textbf{0.8248} &	\textbf{0.8897} \\
	& IMap &	48 &	0.8685 & 0.8784 &	0.8153 &	0.8777	 &  \nores	\\
	& SCDA &	466 &	0.5943 & 0.7550 &	0.2703 &	0.4369	 &  \nores	\\
\midrule
\multirow{5}{*}{COM-E} & &	16 &	0.7128	& 0.7939	& 0.6475	& 0.7444	 & \nores \\
% \multirow{5}{*}{\textbf{0.8971}} &
% \multirow{5}{*}{\textbf{0.9062}} &
% \multirow{5}{*}{\textbf{0.8468}} &
% \multirow{5}{*}{\textbf{0.8991}} \\
 &	& 25 & 0.8415	& 0.8778	& 0.7898	& 0.8492	 & \nores \\
 &	& 38	& \textbf{0.8863}	& \textbf{0.8992}	& \textbf{0.8464}	& \textbf{0.8987} 	 & \textbf{0.8971} & \textbf{0.9062} & \textbf{0.8468} & \textbf{0.8991} \\
 &	& 48 & 0.8727	& 0.8892	& 0.8336	& 0.8861	 & \nores\\
&	& 466 &	0.6088	& 0.7602	& 0.2872	& 0.4529	 & \nores\\
\midrule
\multirow{5}{*}{\CAnv} &	CNM &	16 &	0.7270 &	0.8108 &	0.6471 &	0.7460	 &  0.9218 &	0.9227 &	0.8989 & 0.9187	 \\
 &	LV &	25 &	0.8867 &	0.9232	& 0.8859 &	0.8994	 &  0.9328 &	0.9426 &	0.9078 &	0.9284	 \\
 &	LPA &	38 &	\textbf{0.9371} &	0.9457 &	\textbf{\underline{0.9129}} &	\textbf{\underline{0.9326}}	 &  \textbf{0.9355} &	0.9453 &	0.9097 &	\textbf{\underline{0.9346}}	 \\
 &	IMap &	48 &	0.9283 &	\textbf{0.9488} &	0.9123 &	0.9245	 &  0.9292 &	\textbf{0.9466} &	\textbf{0.9103} & 0.9318	 \\
 &	SCDA  &	466 &	0.6315 &	0.7935 &	0.3489 &	0.6032	 &  0.8986 &	0.9078 &	0.8748 &	0.8997	 \\
\midrule
\multirow{5}{*}{\Mnv} &	CNM &	16 &	0.7260 &	0.8093 &	0.6517 &	0.7521	 &  \nores	 \\
 &	LV  &	25 &	0.8337 &	0.8673 &	0.7839 &	0.8434	 & \nores		 \\ 
 &	LPA & 38	& \textbf{0.8928}	& \textbf{0.9036}	& 0.8427	& 0.8973	 &
 \textbf{0.8943} &	\textbf{0.9047}	& \textbf{0.8460} &	\textbf{0.8966}	 \\ 
 &	IMap &	48	& 0.8912	& 0.9022	& \textbf{0.8443}	& \textbf{0.8977}	 &
 \nores		 \\
 &	SCDA & 466 &	0.6161	& 0.7738	& 0.2853	& 0.4944	 &
 \nores		 \\
\midrule
\multirow{5}{*}{\MCAnv} & CNM & 16	& 0.7289	& 0.8129	& 0.6658	& 0.7548	 & 0.9225 &	0.9264 &	0.8959 & 0.9018	 \\
 &	LV &	25 &	0.8720	& 0.9176	& 0.8769	& 0.8877	 &
0.9331 &	0.9470 &	0.9093 &	0.9338	 \\
 &	LPA &	38	& \textbf{\underline{0.9437}}	& 0.9501	& \textbf{0.9127}	& \textbf{0.9321}	 &
 \textbf{\underline{0.9401}} &	0.9493 &	\textbf{\underline{0.9119}} &	\textbf{0.9337}	 \\
 &	IMap &	48	& 0.9348	& \textbf{\underline{0.9511}} &	0.9119 &	\textbf{0.9321}	 &
  0.9388 & \textbf{\underline{0.9496}} &	0.9114 &	0.9332	 \\
 &	SCDA	&	466 &	0.6435 &	0.7959 &	0.3670 &	0.6111	 &
 0.9091 &	0.9147 &	0.8746 &	0.8968	 \\
\bottomrule
\end{tabular}
%}
\end{center}
\end{table}

%% file: fig-defs/lfr1k.tex
\begin{figure}[t]
 \centering
 \caption{Sensitivity of n2v, \Mnv, \CAnv and \MCAnv~ for different community structures in an LFR network with 1000 nodes (LFR-1K) obtained with the mixing parameter $\mu$ set to $\{0.2, 0.4, 0.6, 0.8\}$.}
 \label{fig:sensitivitylfr1000}
\begin{subfigure}[htb]{.32\textwidth}
    \centering
    \includegraphics[width=\columnwidth]{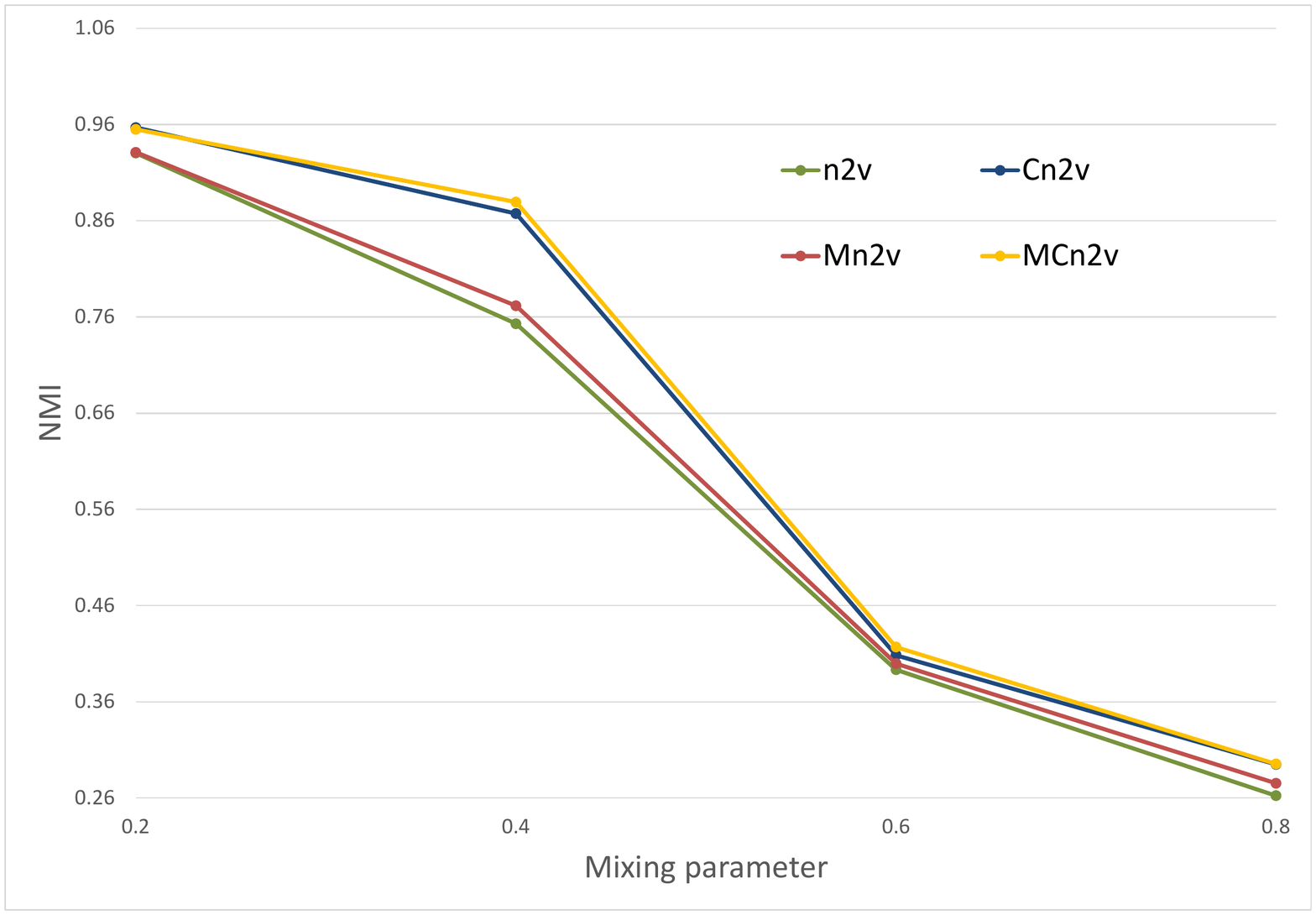}
    \caption{NMI}
    \end{subfigure}
\begin{subfigure}[htb]{.32\textwidth}
    \centering
    \includegraphics[width=\columnwidth]{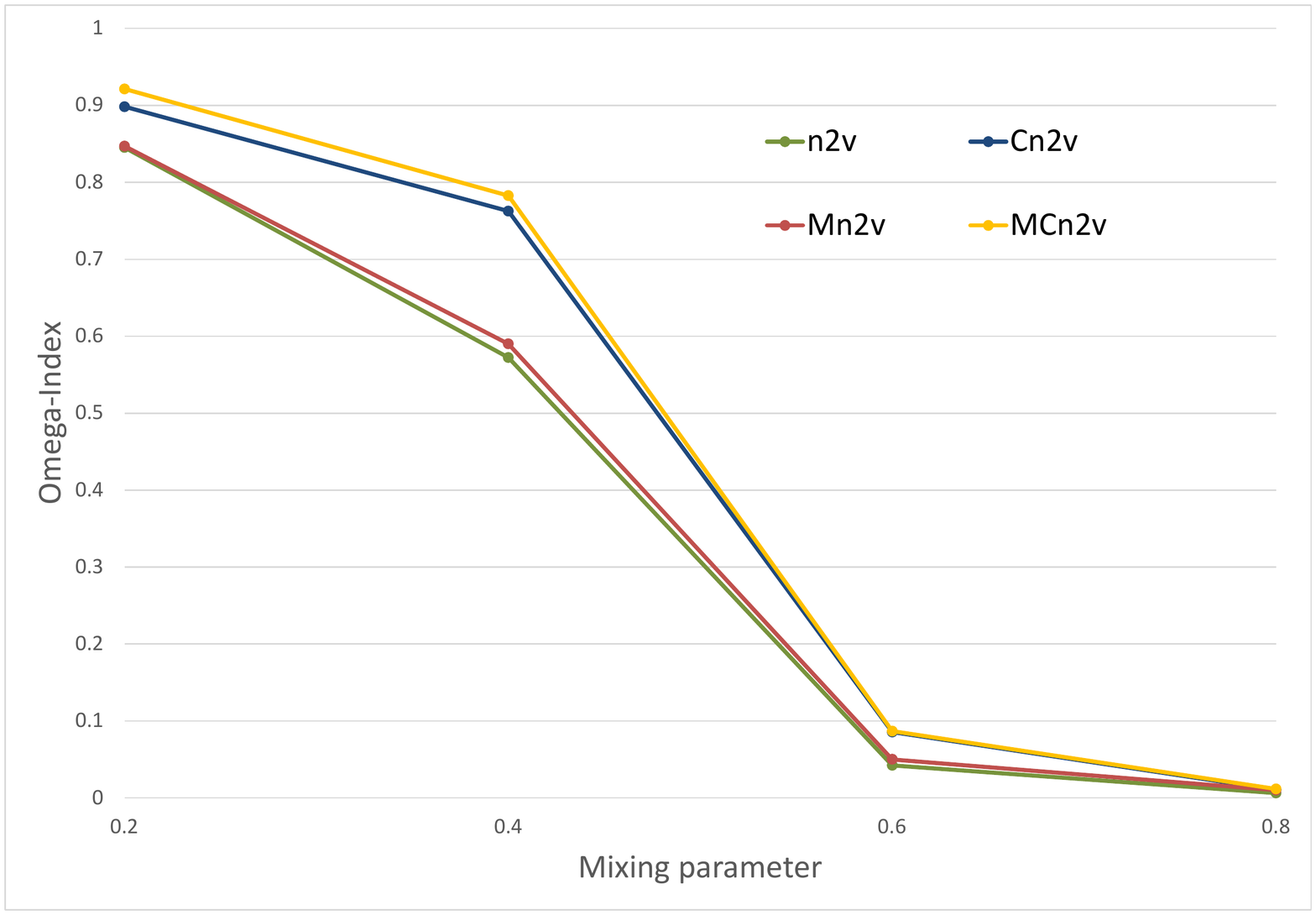}
    \caption{$\Omega$-index }
    \end{subfigure}
\begin{subfigure}[htb]{.32\textwidth}
    \centering
    \includegraphics[width=\columnwidth]{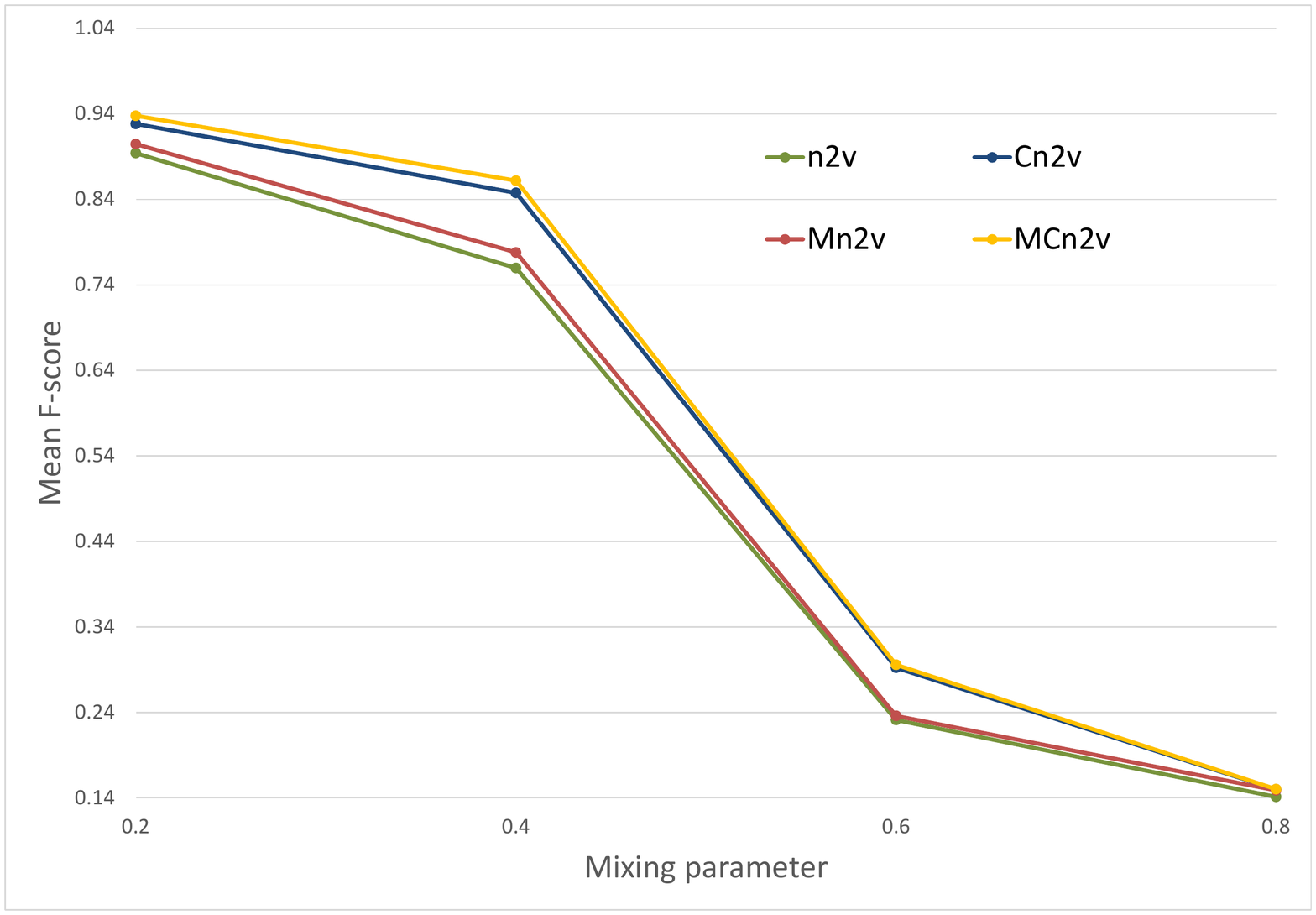}
    \caption{Mean F-score }
    \end{subfigure}
\centering
 \caption{Similar sensitivity analysis as in Figure \ref{fig:sensitivitylfr1000}, on LFR networks with 5000 nodes (LFR-5K).}
 \label{fig:sensitivitylfr5000}
\begin{subfigure}[htb]{.32\textwidth}
    \centering
    \includegraphics[width=\columnwidth]{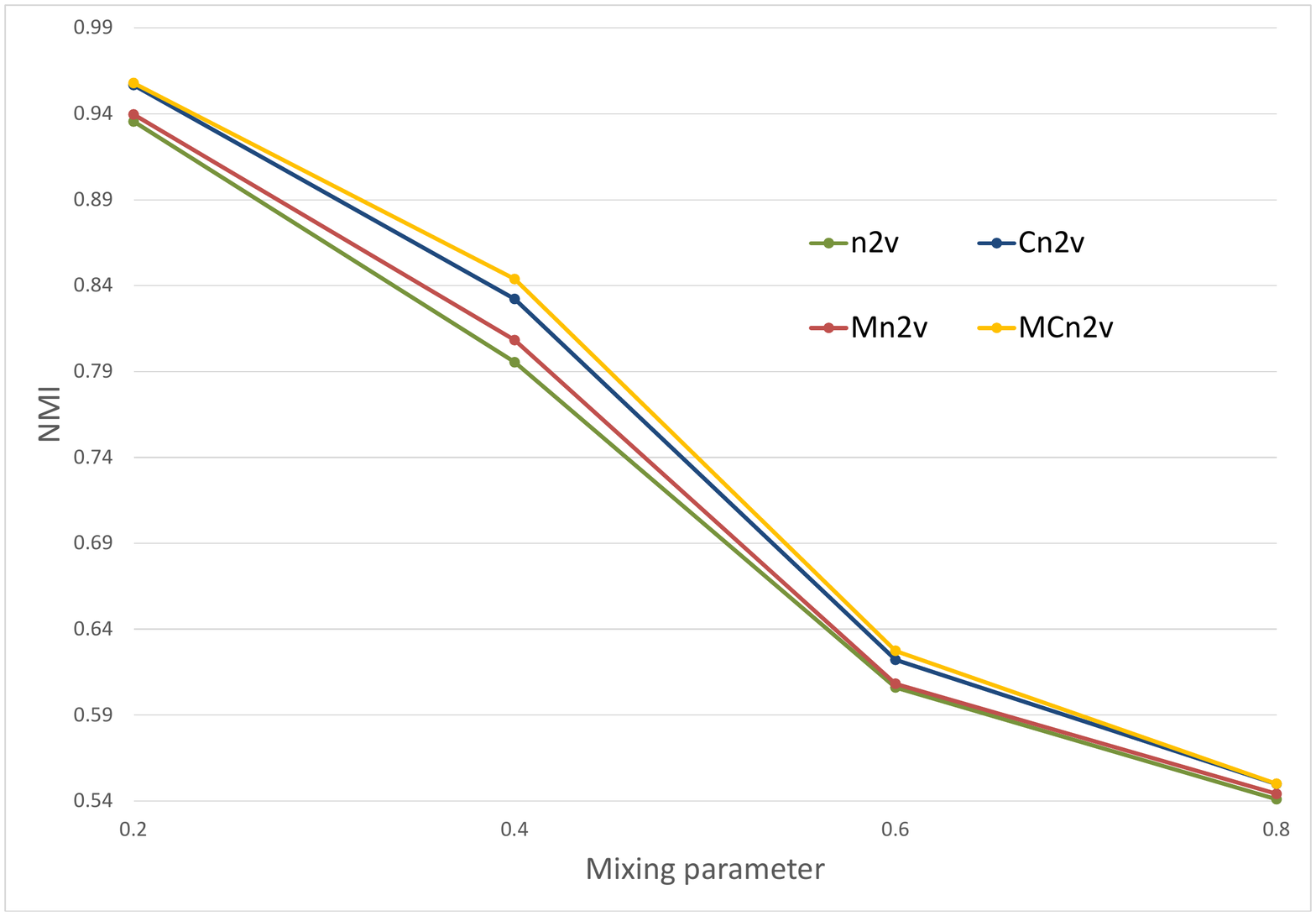}
    \caption{NMI}
    \end{subfigure}
\begin{subfigure}[htb]{.32\textwidth}
    \centering
    \includegraphics[width=\columnwidth]{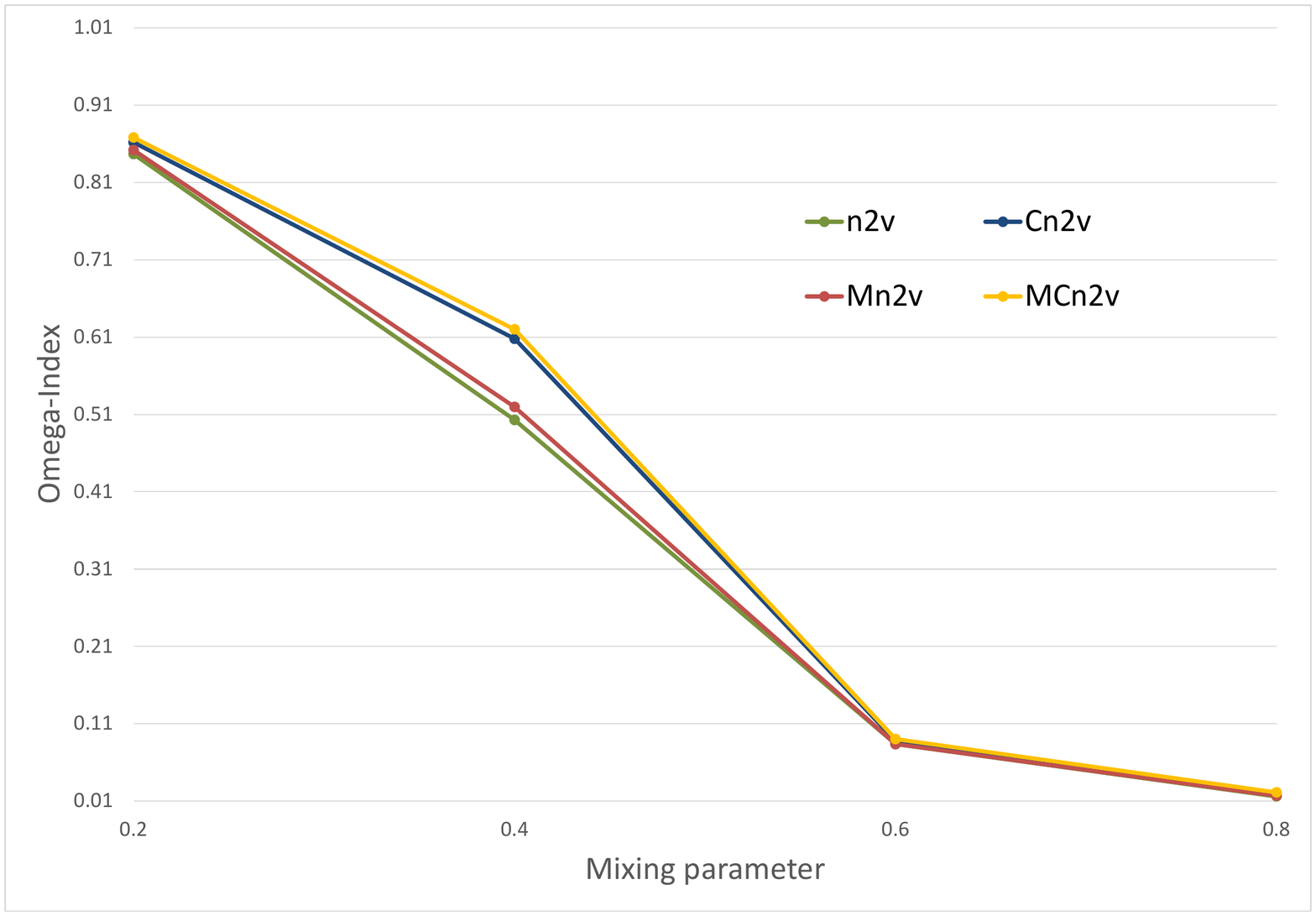}
    \caption{$\Omega$-index }
    \end{subfigure}
\begin{subfigure}[htb]{.32\textwidth}
    \centering
    \includegraphics[width=\columnwidth]{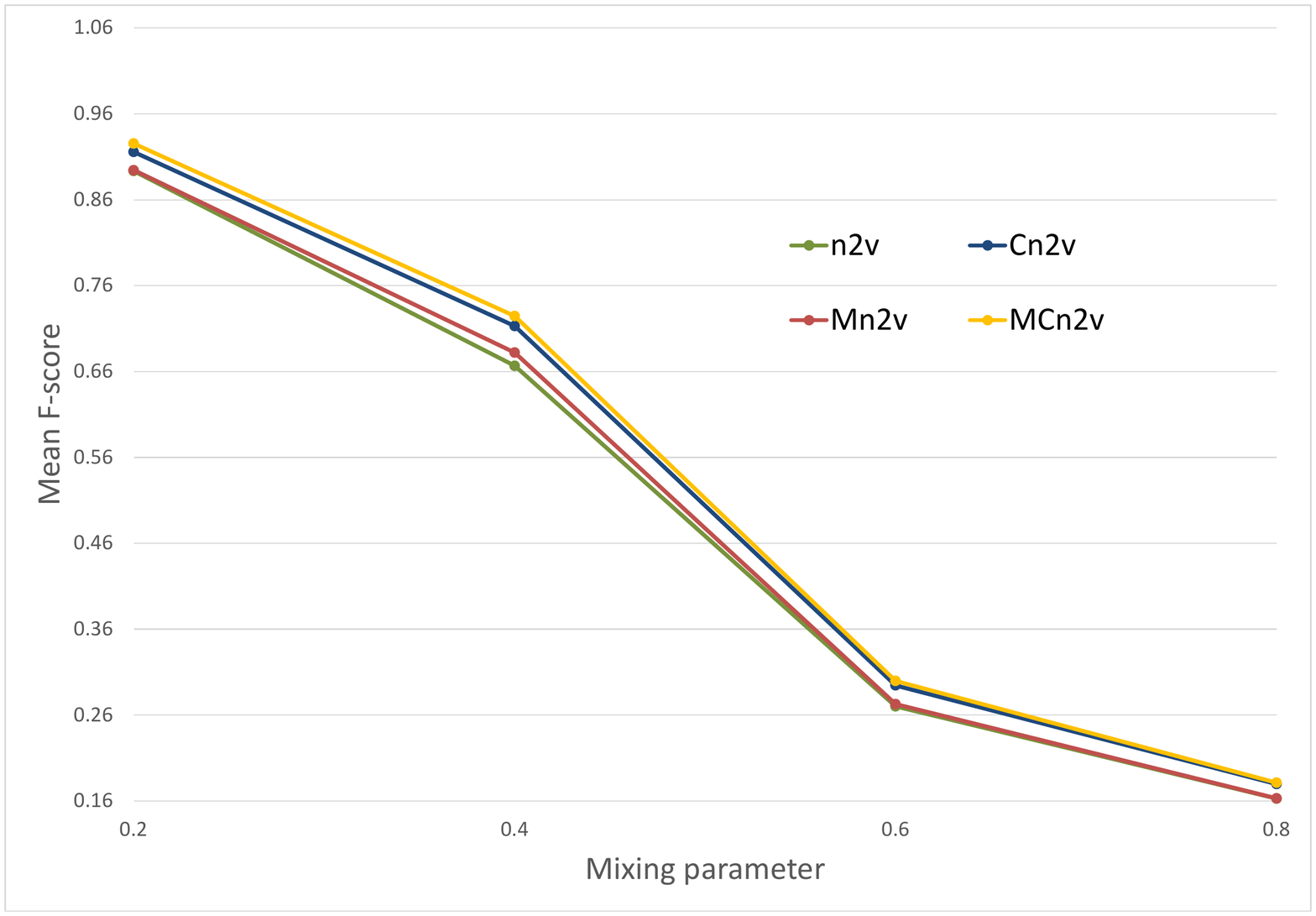}
    \caption{Mean F-score }
    \end{subfigure}
\end{figure}

%% file: fig-defs/mu_2.tex
\begin{figure}[t]
 \centering
 \caption{Sensitivity of \MCAnv~with respect to its parameters $p$, $q$ and window size, $ws$, on LFR-1K network with mixing parameter $\mu=0.2$.}
 \label{fig:sensitivity02}
\begin{subfigure}[t]{.32\textwidth}
    \centering
    \includegraphics[width=\columnwidth]{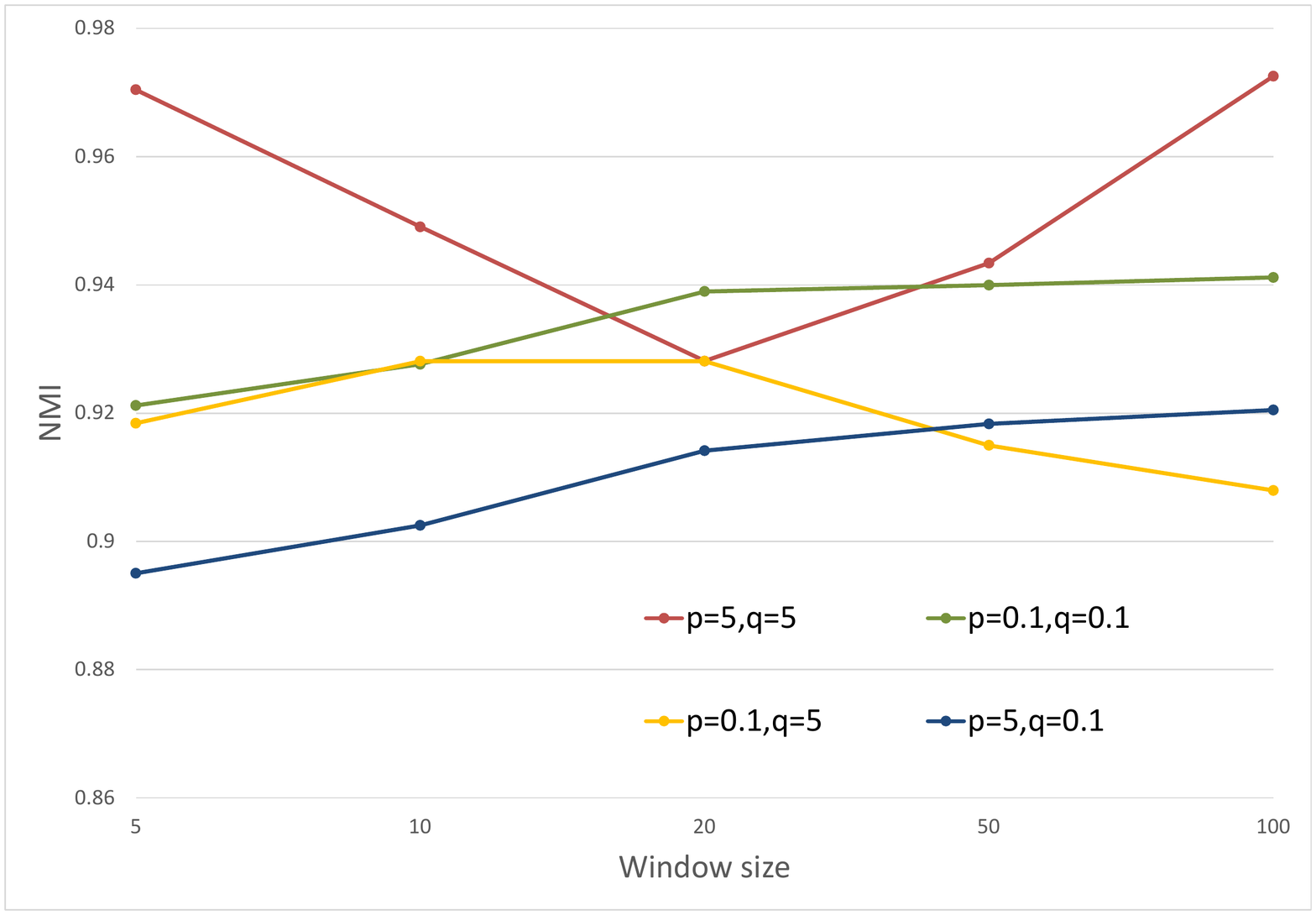}
    \caption{NMI}
    \end{subfigure}
\begin{subfigure}[t]{.32\textwidth}
    \centering
    \includegraphics[width=\columnwidth]{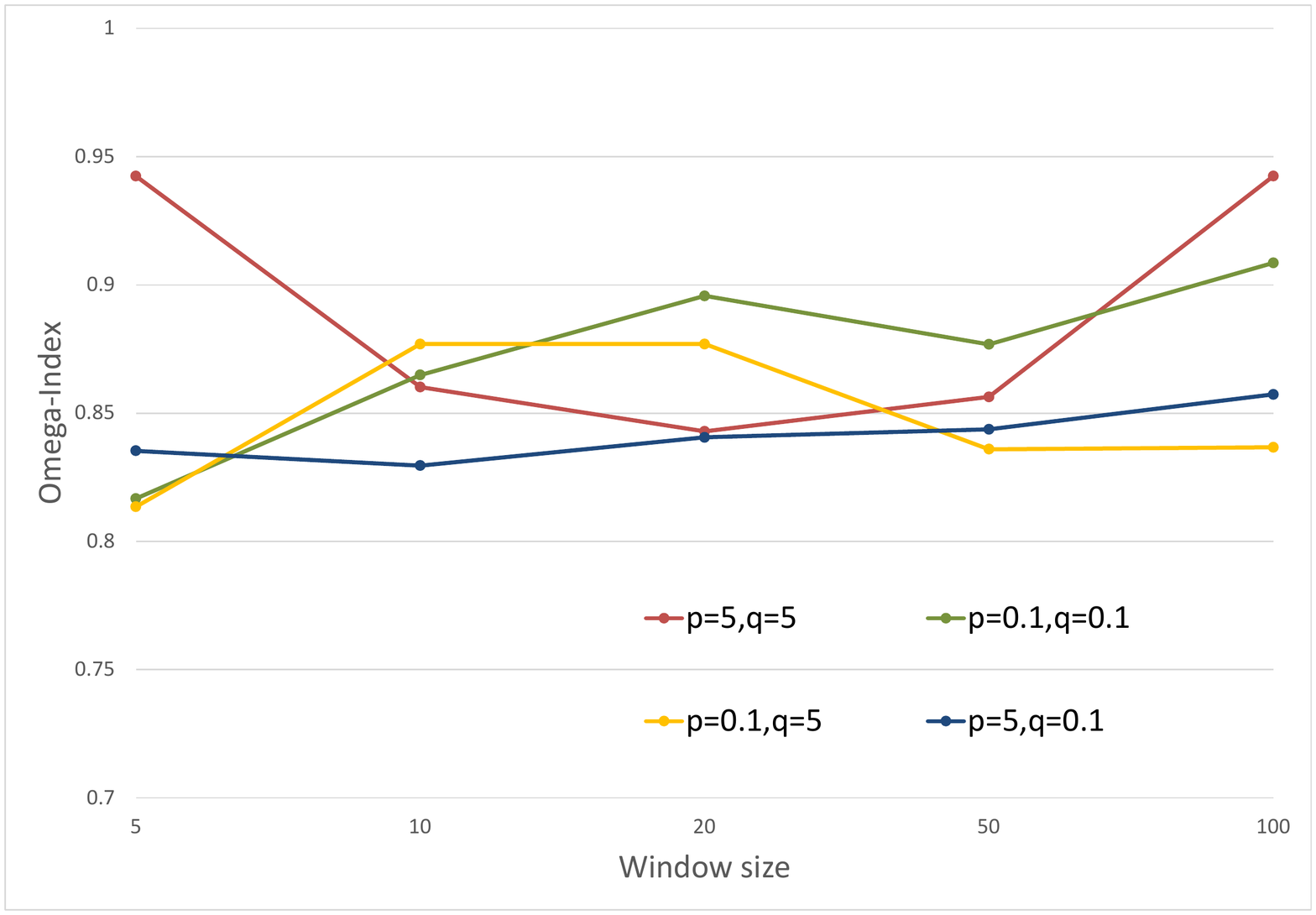}
    \caption{$\Omega$-index }
    \end{subfigure}
\begin{subfigure}[t]{.32\textwidth}
    \centering
    \includegraphics[width=\columnwidth]{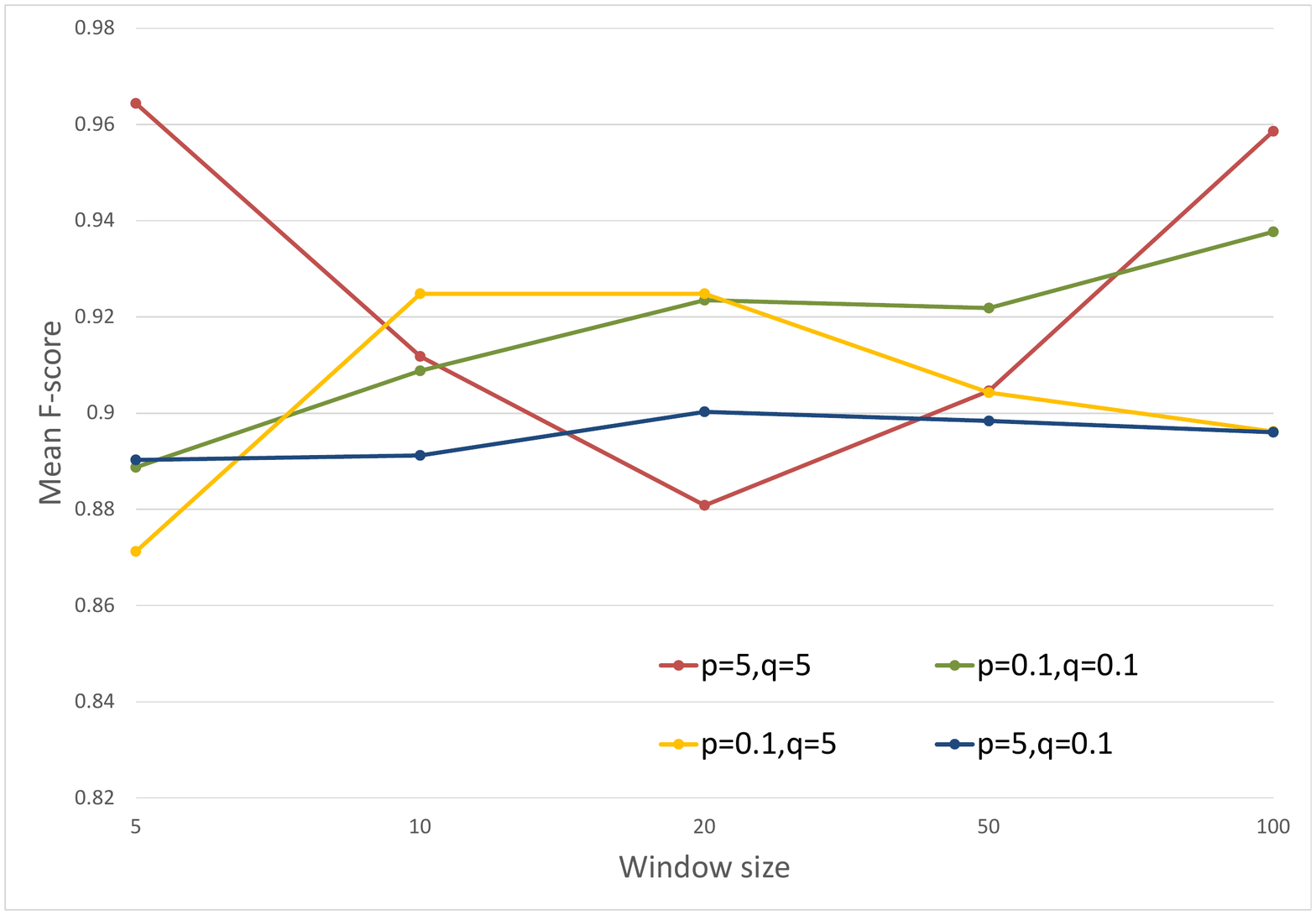}
    \caption{Mean F-score }
    \end{subfigure}
\end{figure}

%% file: fig-defs/mu_4.tex
\begin{figure}[t]
 \centering
 \caption{Sensitivity of \MCAnv~with respect to its parameters $p$, $q$ and window size, $ws$, on LFR-1K network with mixing parameter $\mu=0.4$.}
 \label{fig:sensitivity04}
\begin{subfigure}[t]{.32\textwidth}
    \centering
    \includegraphics[width=\columnwidth]{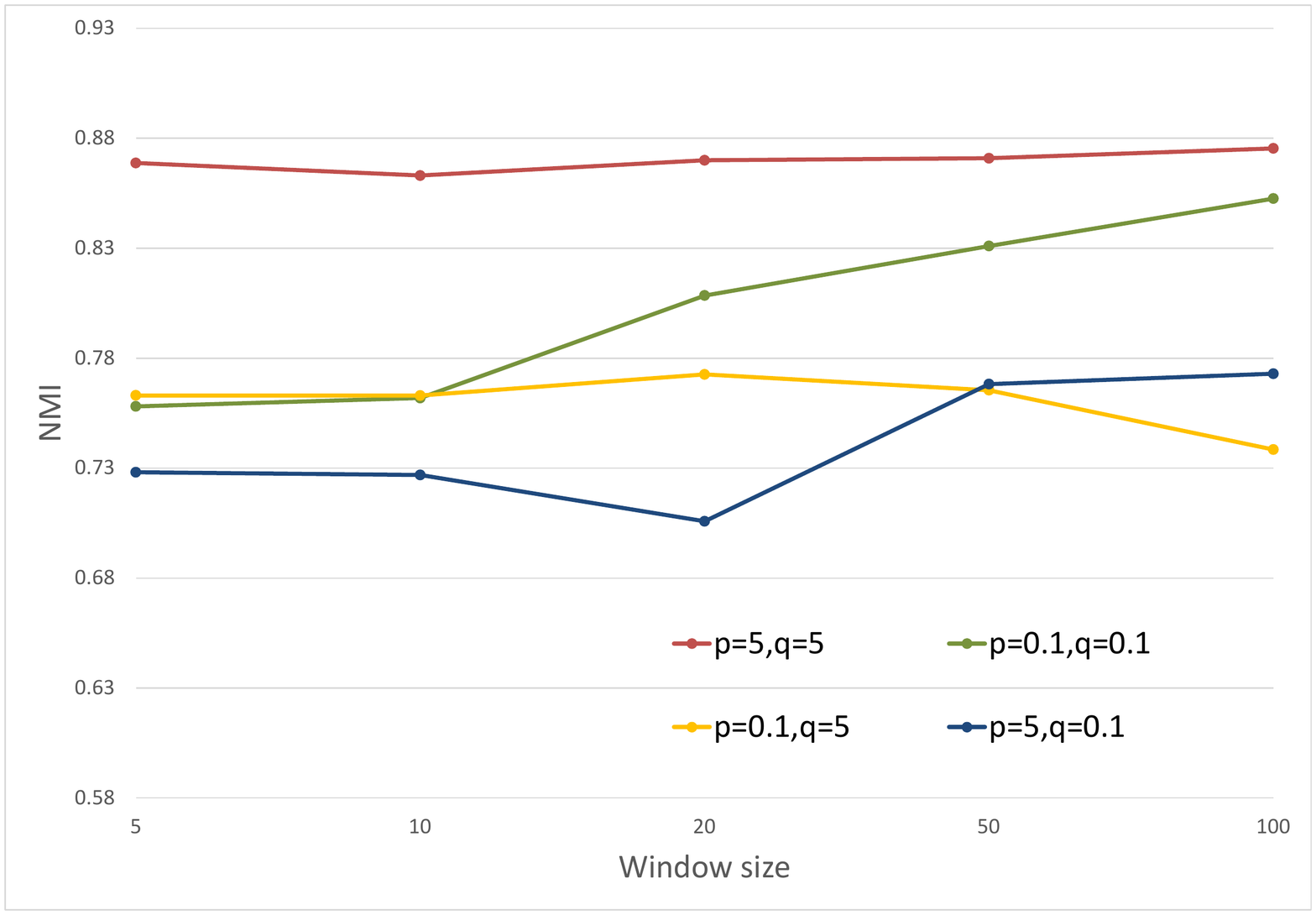}
    \caption{NMI}
    \end{subfigure}
\begin{subfigure}[t]{.32\textwidth}
    \centering
    \includegraphics[width=\columnwidth]{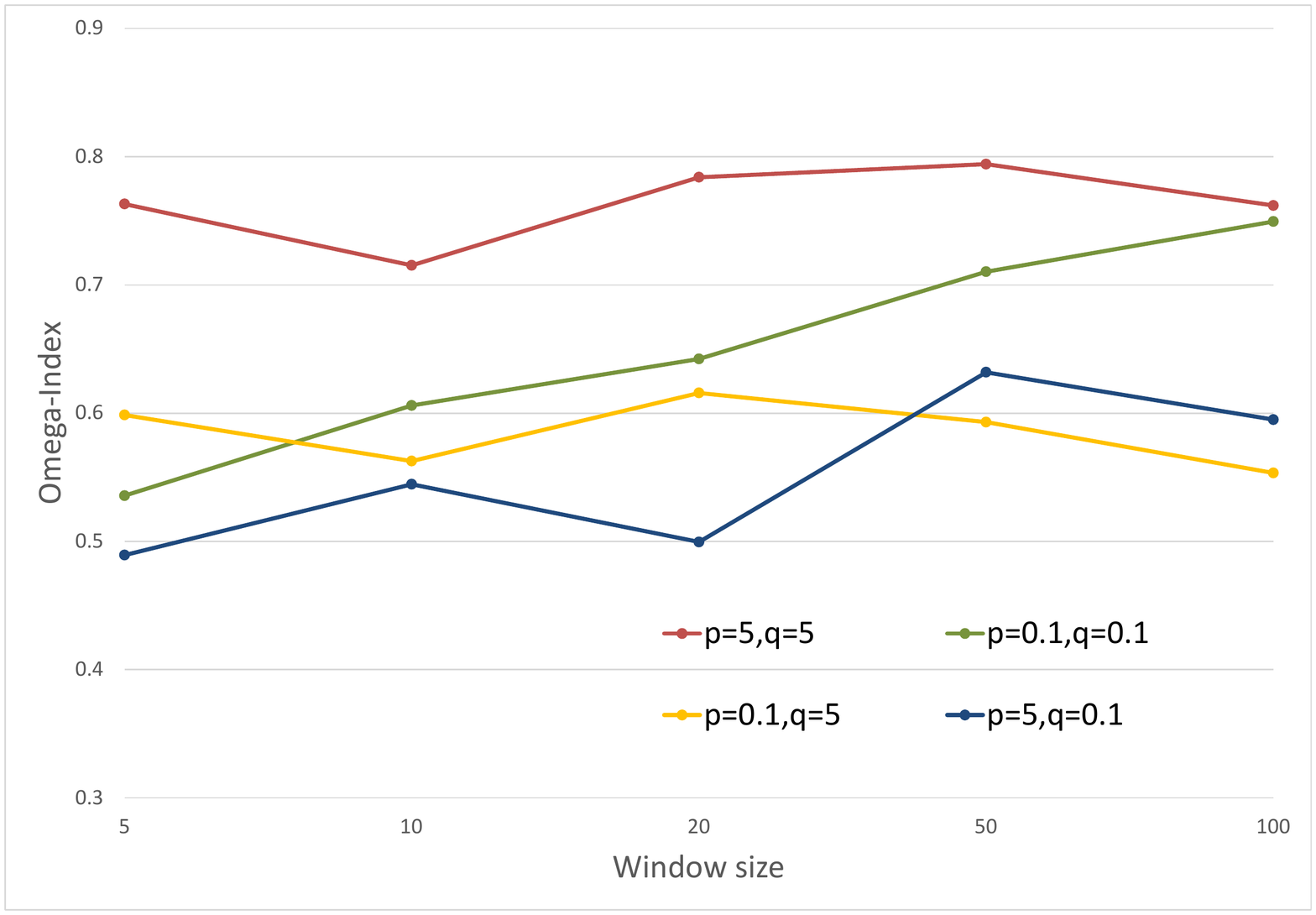}
    \caption{$\Omega$-index }
    \end{subfigure}
\begin{subfigure}[t]{.32\textwidth}
    \centering
    \includegraphics[width=\columnwidth]{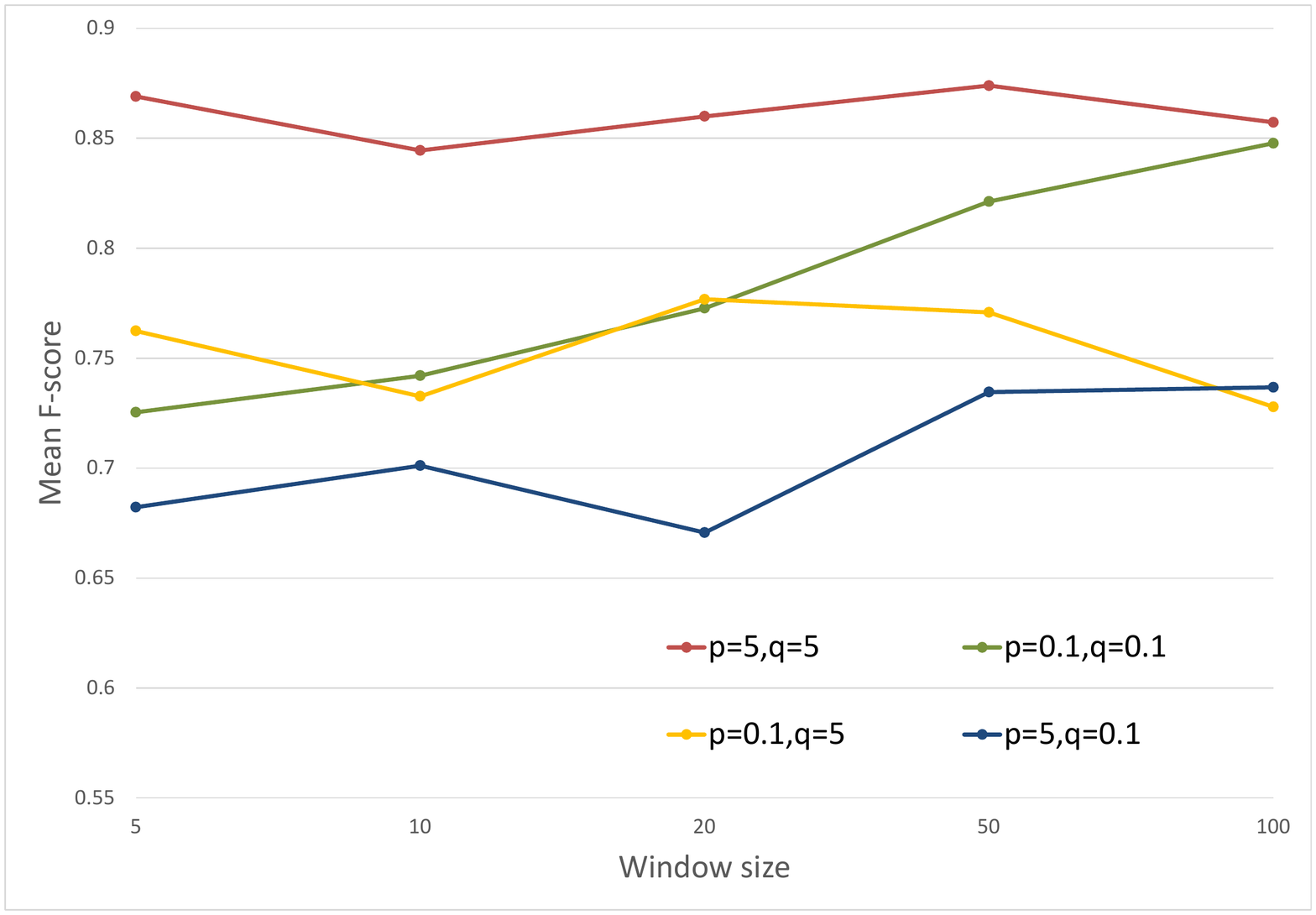}
    \caption{Mean F-score }
    \end{subfigure}
\end{figure}

%% file: fig-defs/mu_6.tex
\begin{figure}[t]
 \centering
 \caption{Sensitivity of \MCAnv~with respect to its parameters $p$, $q$ and window size, $ws$, on LFR-1K network with mixing parameter $\mu=0.6$.}
 \label{fig:sensitivity06}
\begin{subfigure}[t]{.32\textwidth}
    \centering
    \includegraphics[width=\columnwidth]{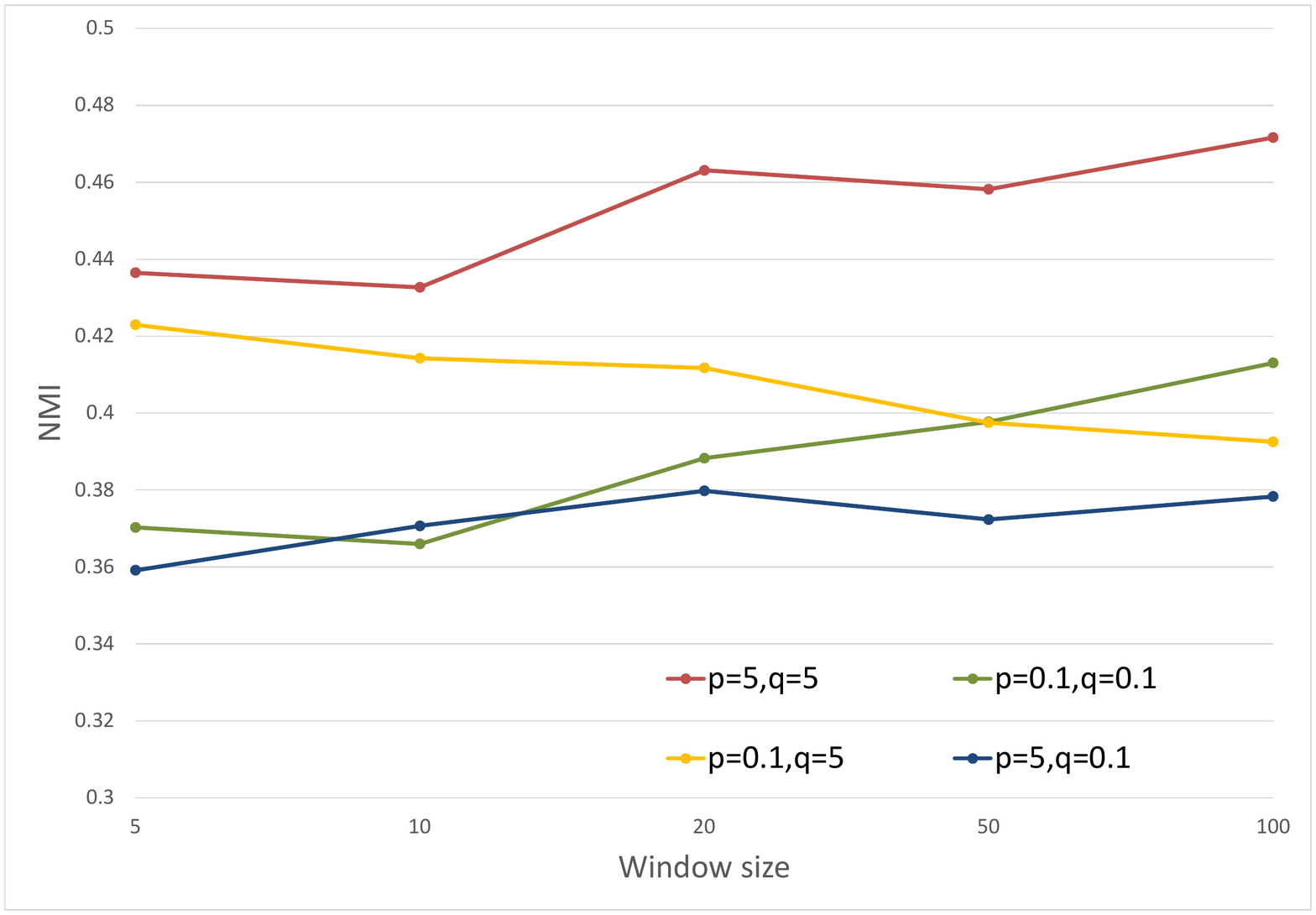}
    \caption{NMI}
    \end{subfigure}
\begin{subfigure}[t]{.32\textwidth}
    \centering
    \includegraphics[width=\columnwidth]{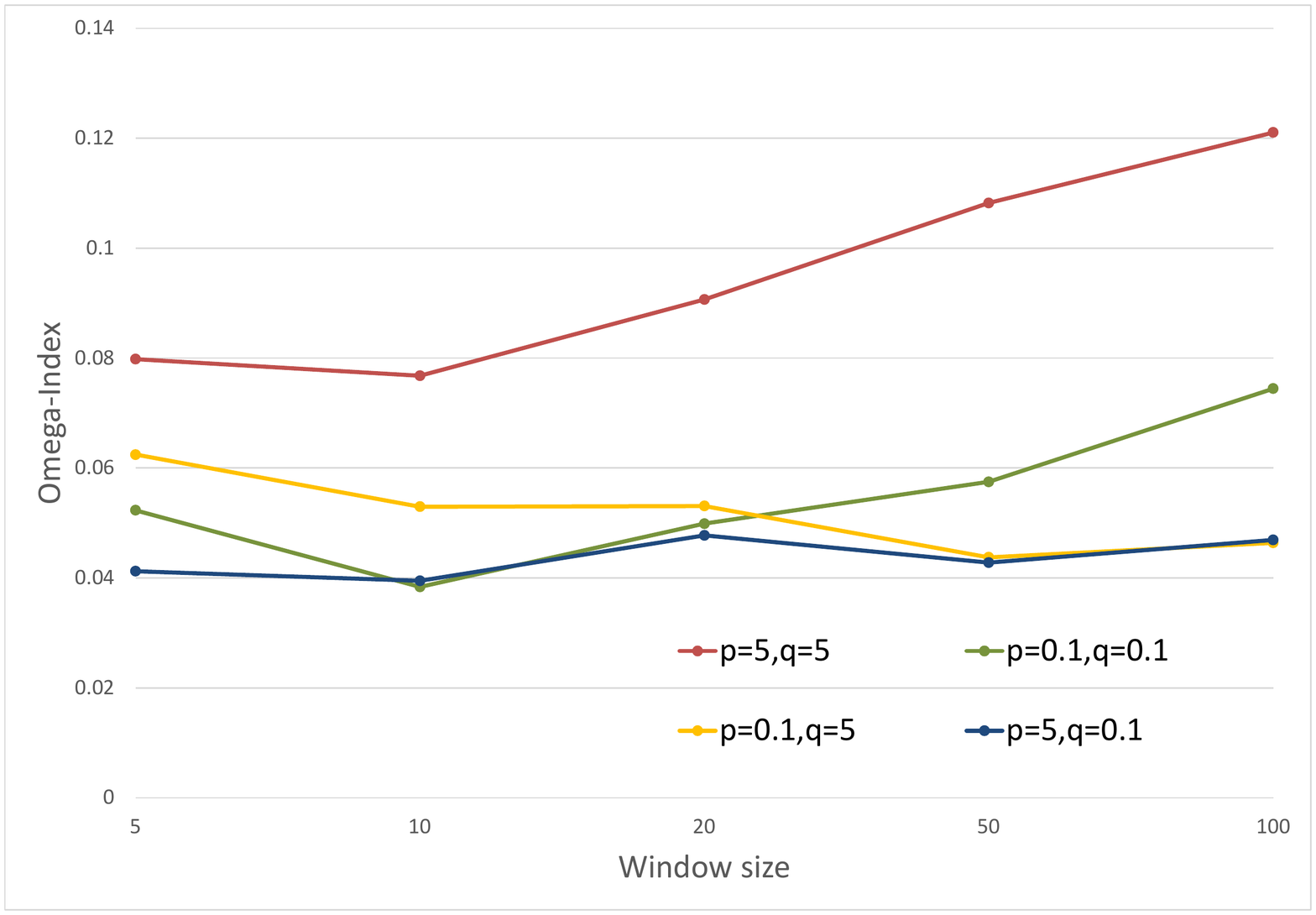}
    \caption{$\Omega$-index }
    \end{subfigure}
\begin{subfigure}[t]{.32\textwidth}
    \centering
    \includegraphics[width=\columnwidth]{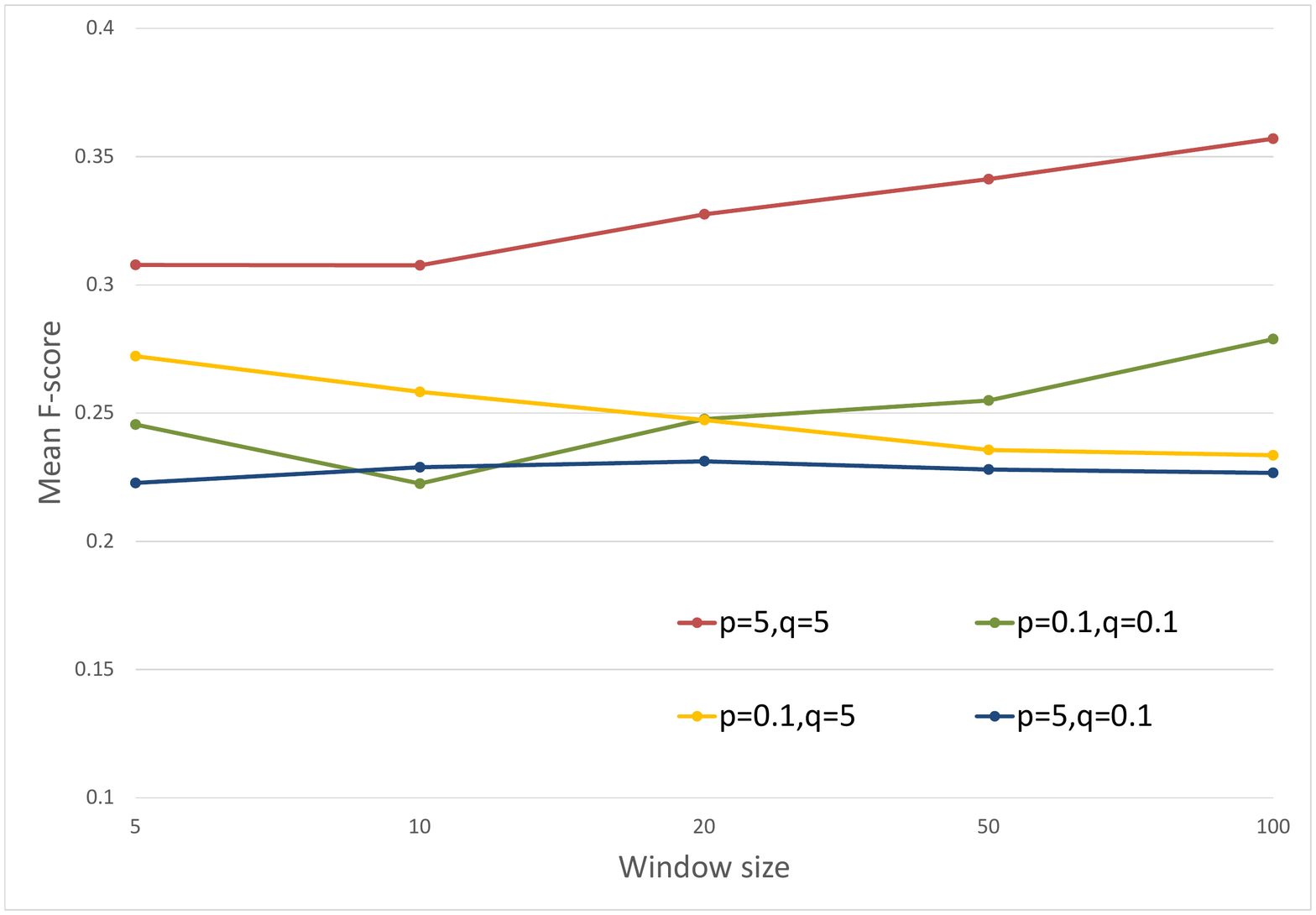}
    \caption{Mean F-score }
    \end{subfigure}
\end{figure}

%% file: fig-defs/mu_8.tex
\begin{figure}[t]
 \centering
 \caption{Sensitivity of \MCAnv~with respect to its parameters $p$, $q$ and window size, $ws$, on LFR-1K network with mixing parameter $\mu=0.8$.}
 \label{fig:sensitivity08}
\begin{subfigure}[t]{.32\textwidth}
    \centering
    \includegraphics[width=\columnwidth]{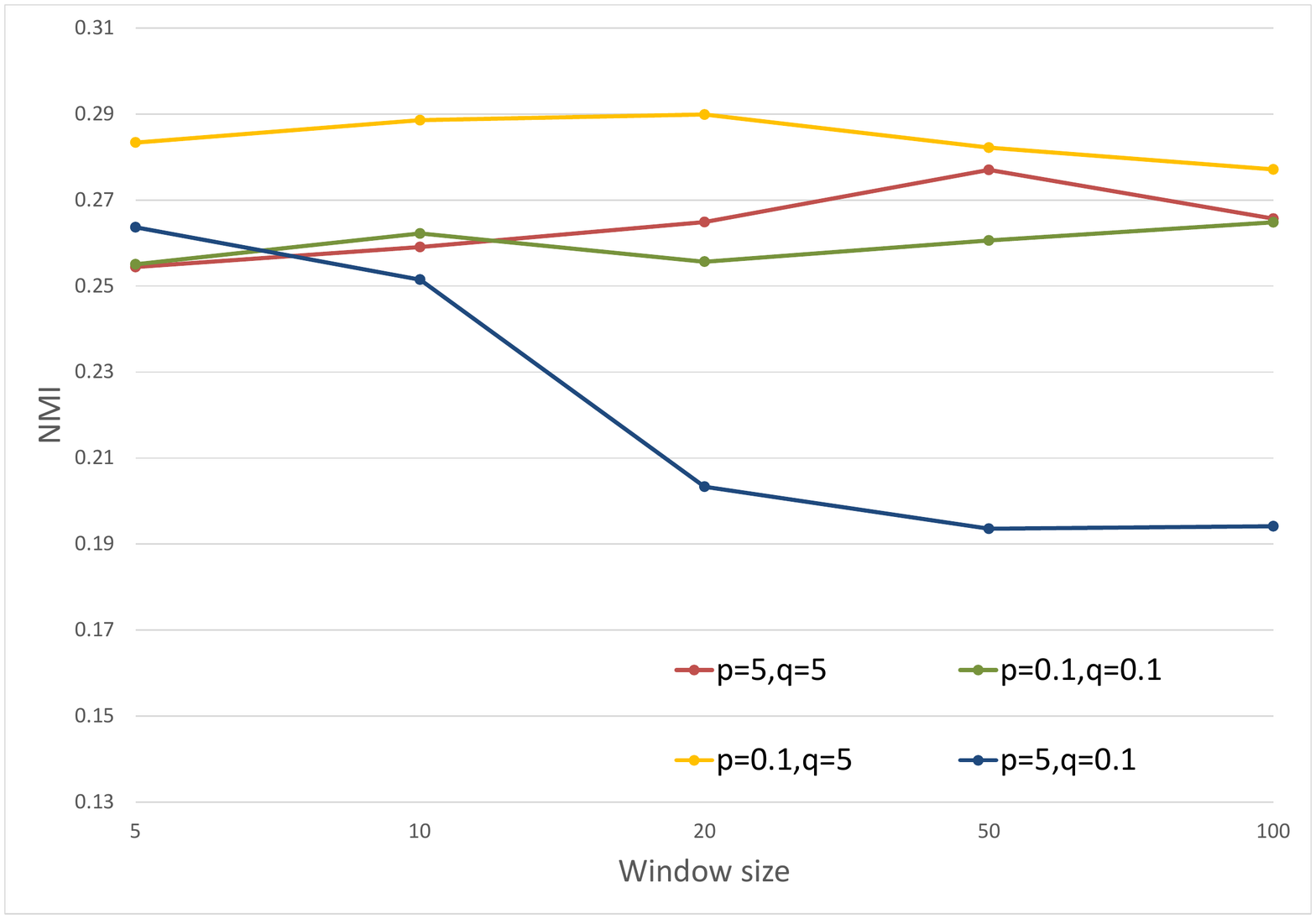}
    \caption{NMI}
    \end{subfigure}
\begin{subfigure}[t]{.32\textwidth}
    \centering
    \includegraphics[width=\columnwidth]{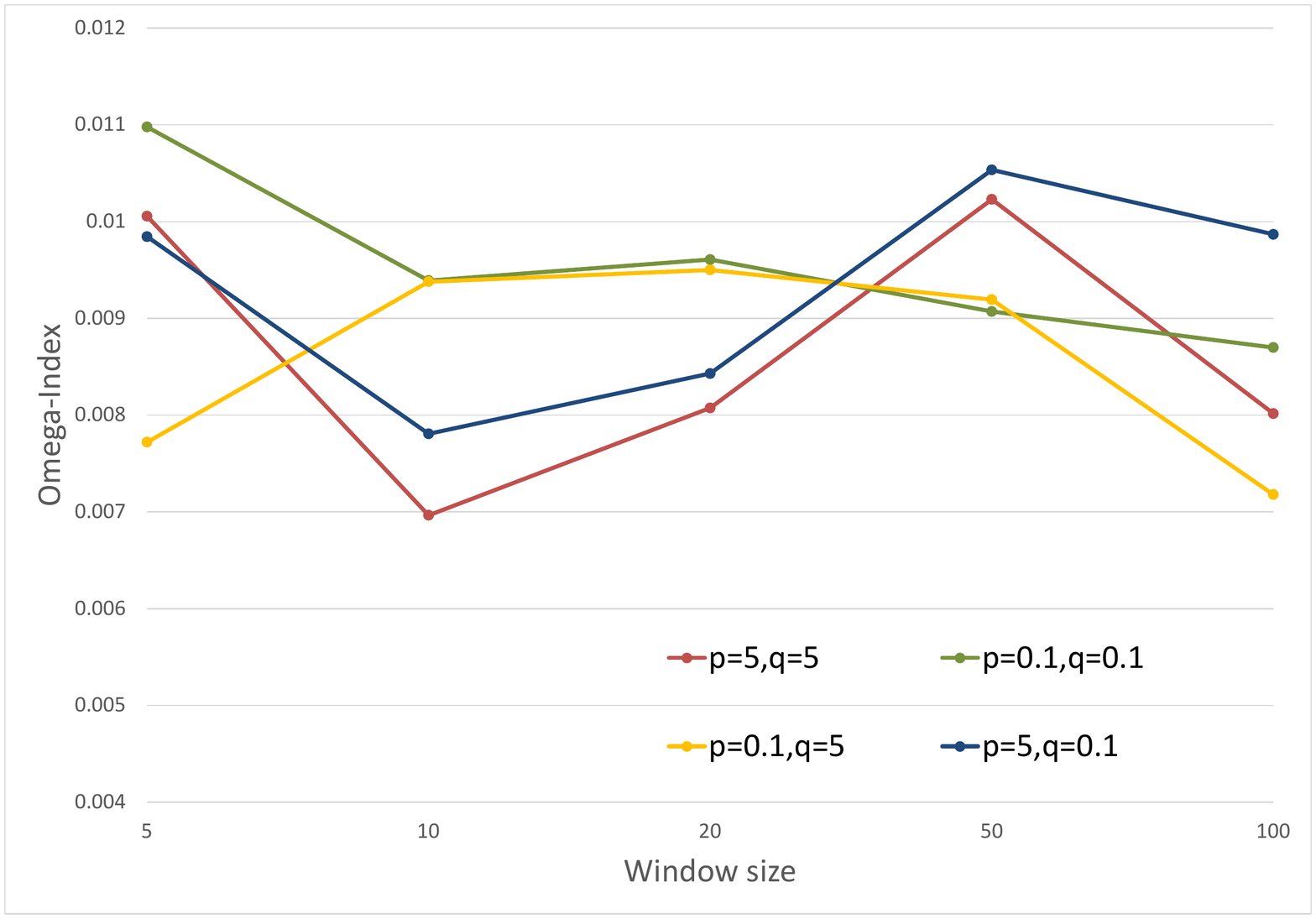}
    \caption{$\Omega$-index }
    \end{subfigure}
\begin{subfigure}[t]{.32\textwidth}
    \centering
    \includegraphics[width=\columnwidth]{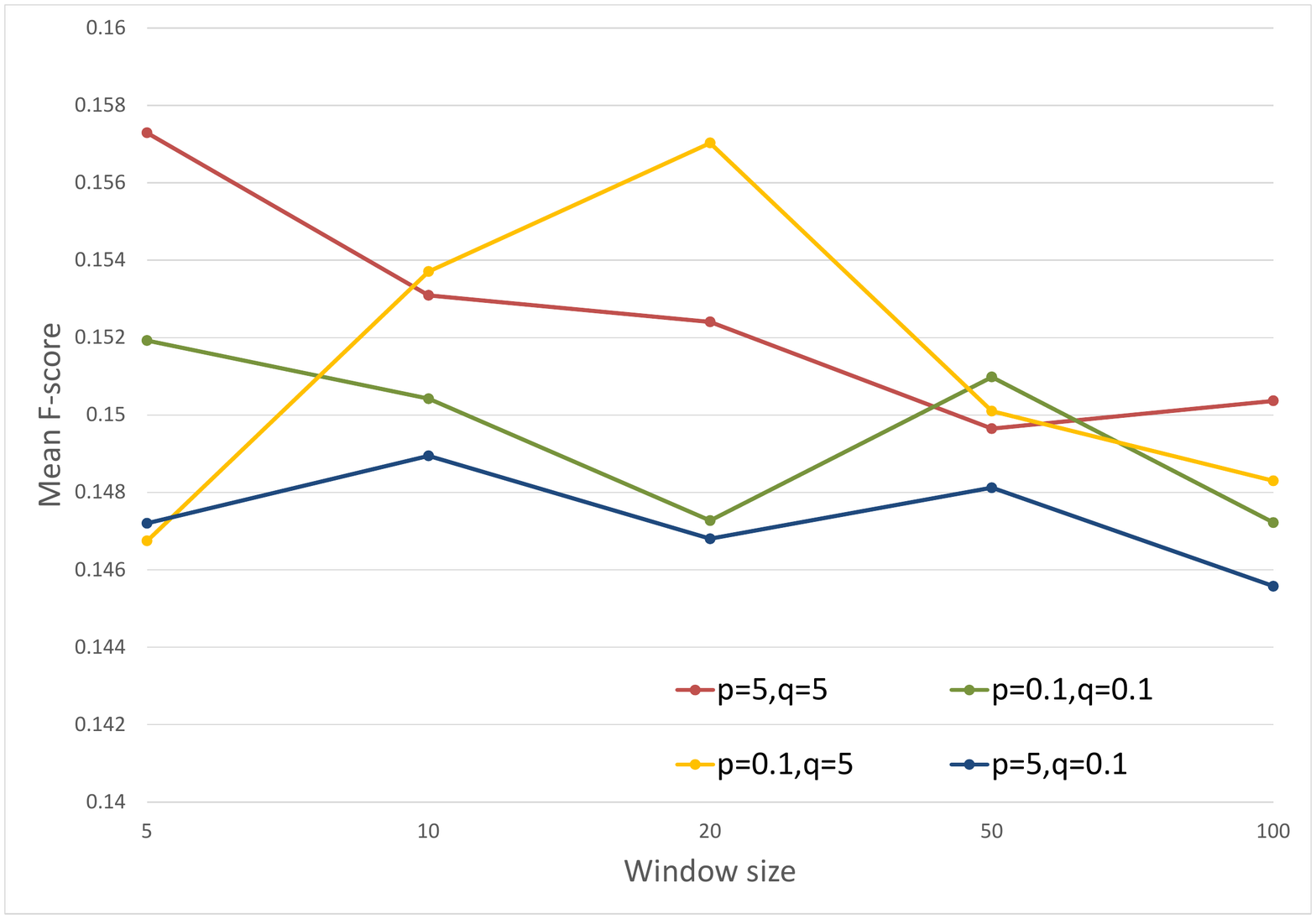}
    \caption{Mean F-score }
    \end{subfigure}
\end{figure}